\documentclass[aps,prb,twocolumn,superscriptaddress,showpacs,showkeys,amsmath,amssymb]{revtex4-1}
\usepackage{amsfonts}
\usepackage{amssymb,amsmath}
\usepackage{graphicx}
\usepackage{bm}
\usepackage{color}

\newcommand{\newc}{\newcommand}
\newc{\beq}{\begin{equation}}
\newc{\eeq}{\end{equation}}
\newc{\beqa}{\begin{eqnarray}}
\newc{\eeqa}{\end{eqnarray}}
\newc{\mbf}{\mathbf}
\newc{\dg}{\dagger}
\newc{\rkilj}{{{\bf R}^{ij}_{kl}}}
\newc{\rijabv}{{\mbf{r}}_{ij}^{\alpha \beta}}
\newc{\rijab}{{{r}}_{ij}^{\alpha \beta}}
\newc{\Ski}{{\mbf{{S}}_{k i}}}
\newc{\Skica}{{\text{S}}{^{\alpha}_{ki}}}
\newc{\Sljcb}{{\text{S}}{^{\beta}_{lj}}}
\newc{\Sqa}{{\mbf{{S}}_{q \alpha}}}
\newc{\Slj}{{\mbf{{S}}_{l j}}}
\newc{\Sjbcm}{{\text{S}}{^m_{j \beta}}}
\newc{\Sjbcn}{{\text{S}}{^n_{j \beta}}}
\newc{\Sqica}{{\text{S}}{^{\alpha}_{i}}({\bf{k}})}
\newc{\Smqica}{{\text{S}}{^{\alpha}_{i}}(-{\bf{k}})}
\newc{\Sqjcb}{{\text{S}}{^{\beta}_{j}}({\bf{k}})}
\newc{\Smqjcb}{{\text{S}}{^{\beta}_{j}}(-{\bf{k}})}
\newc{\eia}{{\hat{\mbf{e}}}_{i \alpha}}
\newc{\eib}{{\hat{\mbf{e}}}_{i \beta}}
\newc{\ejb}{{\hat{\mbf{e}}}_{i \beta}}
\newc{\eig}{{\hat{\mbf{e}}}_{i \gamma}}
\newc{\ebm}{{\hat{\mbf{e}}}_{\beta m}}
\newc{\ebn}{{\hat{\mbf{e}}}_{\beta n}}
\newc{\modrijab}{\lvert {\mbf{r}}_{ij}^{\alpha \beta} \rvert}

\begin{document}

\title{Classical dipoles on the kagome lattice}

\author{Mykola Maksymenko}
\affiliation{Max-Planck-Institut f\"{u}r Physik komplexer Systeme,
        N\"{o}thnitzer Stra{\ss}e 38, 01187 Dresden, Germany}
\affiliation{Department of Condensed Matter Physics, Weizmann Institute of Science, Rehovot 76100, Israel}
\email{E-mail: mykola.maksymenko@weizmann.ac.il}
\author{V. Ravi Chandra}
\affiliation{School of Physical Sciences, National Institute of Science Education and Research, Institute of Physics Campus, Bhubaneswar, 751005 India}
\author{Roderich Moessner}
\affiliation{Max-Planck-Institut f\"{u}r Physik komplexer Systeme,
        N\"{o}thnitzer Stra{\ss}e 38, 01187 Dresden, Germany}



\begin{abstract}
Motivated by recent developments in magnetic materials, frustrated nanoarrays and cold atomic systems, 
we investigate the behaviour of dipolar spins on the frustrated two-dimensional kagome lattice. By combining the Luttinger-Tisza approach, numerical energy minimization, spin-wave analysis and parallel tempering Monte-Carlo, we study long-range ordering and finite-temperature phase transitions for a Hamiltonian containing both dipolar and nearest-neighbor interactions. For both weak and moderate dipolar interactions, the system enters a three-sublattice 
long-range ordered state, with each triangle having vanishing dipole and quadrupole moments; 
while for dominating dipolar interactions we uncover ferrimagnetic three-sublattice order. These are also
the ground states for XY spins. 
We discuss excitations of, as well as phase transitions into, these states. We find behaviour consistent
with Ising criticality for the $120^o$ state, while the ferrimagnetic state appears to be associated with drifting exponents. The celebrated flat band of zero-energy excitations of the kagome 
nearest-neighbour Heisenberg model 
is lifted to finite energies but acquires only minimal dispersion as dipolar interactions are added.
\end{abstract}

\maketitle

\section{Introduction}
Long ranged dipolar interactions occur in any lattice system of interacting magnetic
moments. 
However, the assessment of the relevance of  dipolar interactions in determining the 
behavior of magnetic systems has witnessed 
a recalibration in the recent past. 
This is largely due to the advent 
of several experimental systems that shifted the focus away from purely exchange coupled 
magnets where the dipolar interaction is routinely neglected.

We can identify at least three broad classes of systems which have led to this renewed interest in 
dipolar interactions. The first are the $A_2 B_2 O_7$ pyrochlore oxides, which  
most closely resemble conventionally studied magnetic systems \cite{gardner_gingras_greedan_rmp}. 
For these, as a result of an interplay of crystal field effects, geometry and
the specific magnetic ions involved, 
the dipolar interations can be appreciable. 
A second class are nanomagnetic arrays \cite{nisoli_moessner_schiffer_rmp},
collections of nanomagnetic islands arranged in a regular pattern 
using lithography. The magnitude of the moments as well as the strength of the dipolar interactions
can be tuned to a great degree by controlling the dimensions and separation of the magnetic islands. These systems 
are much more tunable than the thin film systems studied in the past with a view to analysing pattern formation and ordering
via the dipolar interaction \cite{de_bell_review_thin_films}. Finally, the last decade has seen rapid development of 
magnetic systems of polar molecules and atomic gases with large dipole moments confined in optical lattices \cite{pupilo_et_al_cold_atoms_review, peter2012anomalous}. 

Of particular interest is the interplay of dipolar interactions and geometrical frustration. 
On frustrated lattices, an  exchange term typically gives rise to a macroscopically degenerate
yet locally strongly constrained ground state manifold, usually lacking conventional
magnetic order. This constraint can be thought of as restricting the space of states, often in a 
topologically non-trivial way, within which dipolar interactions are to be minimised; or, conversely,
the dipolar interactions can be thought of as lifting the degeneracy, akin to the usual order-by-disorder
physics characteristic of quantum and thermal fluctuations \cite{moessner2001magnets}. The combination of exchange and dipoles can lead to suprising results, such
as in the case of spin ice \cite{bramwell2001spin}, where the underlying elementary 
excitations can be seen as doubly 
gauge charged\cite{moessner2010irrational}
(emergent) magnetic monopoles \cite{spin_ice_monopoles_nature}.

Theoretical efforts to study dipolar spins are several decades old \cite{luttinger_tisza, brankov_danchev_2d_arbitrary_angle, rozenbaum1982phase,maleev_1976}. An early milestone 
 is the work of Luttinger and Tisza \cite{luttinger_tisza} who established that the ground state
for a simple cubic lattice of dipoles is an antiferromagnetic arrangement of chains of aligned dipoles. 
Thereafter, Maleev \cite{maleev_1976} found that the long range and anisotropic nature of the dipolar interactions 
can stabilise long range order in two dimensional magnets - 
something that is prohibited for short ranged isotropic exchange Hamiltonians 
because of the Mermin-Wagner theorem. 
Indeed, for nanomagnetic arrays and cold atoms in optical lattices the study of 
two dimensions is particularly relevant. 
For dipoles on the square lattice the ground state likely consists of antiferromagnetically aligned 
ferromagnetic legs \cite{prakash_henley, fernandez_alonso_square_lattice_2007} 
and closely related degenerate states \cite{belorobov19832Ddipolar}. 
For the triangular lattice a ferromagnetic phase has been reported for purely dipolar
interactions but it was argued that other phases like a $120^{o}$ phase and striped 
antiferromagnetic phases appear for increasing strength of the exchange interaction \cite{rastelli_regina_tassi_triangular_2007, politi2006dipolar,danchev1990spherical,sasaki1998_MC_dipolar_triangular}. 
While there is some agreement about the nature of the low temperature phase for several of these systems
the precise details of the transition to those low temperature phases are frequently under debate. The principal 
reason lies in the subtleties involved in the thermodynamic limit in the presence of
long ranged (and anisotropic) interactions.

For some otherwise well-studied lattices, not even the dipolar ground state is known. 
A case in point are classical dipolar spins on the kagome lattice, the focus
of this work. The  kagome is perhaps the most-studied two-dimensional highly frustrated lattice,
for which even the low-temperature behaviour of a 'simple' nearest-neighbour Heisenberg model 
is remarkably intricate \cite{white2011kagome_science, chern2013kagome} 
We investigate in this paper using a combination of Luttinger-Tisza (LT) method,
spinwave calculations, numerical energy minimisation 
and extensive Monte Carlo simulations the interplay of 
exchange and dipolar
interactions. We find two distinct low-temperature orderings. 
For weak dipolar interactions we observe  $120^{o}$ three-sublattice order with zero net moment;
while for strong dipolar interactions we find a peculiar ferrimagnetic state with continuously varying
net moment. 
Thus we have two different three-sublattice $\mbf{k}_0=(0,0)$ states
at weak and strong dipolar interactions (Fig.~\ref{lattice}b). While our results for the case 
of strong dipolar interactions predict a finite moment per unit cell as in earlier work
\cite{tomita_kagome}, 
our extensive simulations and analytic considerations do not support the existence 
of a disordered non-magnetic sublattice. 

The outline of the paper is as follows. In Section \ref{sec:model} we introduce model and  
 conventions used. In Section \ref{sec:luttinger_tisza} we present the ground-state phase diagram from
  the Luttinger-Tisza approach \cite{luttinger_tisza}. This method 
fails in the case of strong dipolar term and hence in Section \ref{sec:numerical_minimization} we perform a numerical search 
for the ground state and in Section \ref{sec:spin_waves} we confirm that this state is locally stable via
a spinwave analysis. Finally, in Section \ref{section_mc_simulation} we analyze our model using a parallel-tempering 
Monte-Carlo method which confirms our predictions for the ground states and provides the details of 
the corresponding rich finite-temperature  phase transitions. We close with a discussion section. 

\section{Model}
\label{sec:model}
\begin{figure}
\centering
\includegraphics[trim=3cm 10cm 9.4cm 0cm, clip=true, width=0.9\columnwidth]{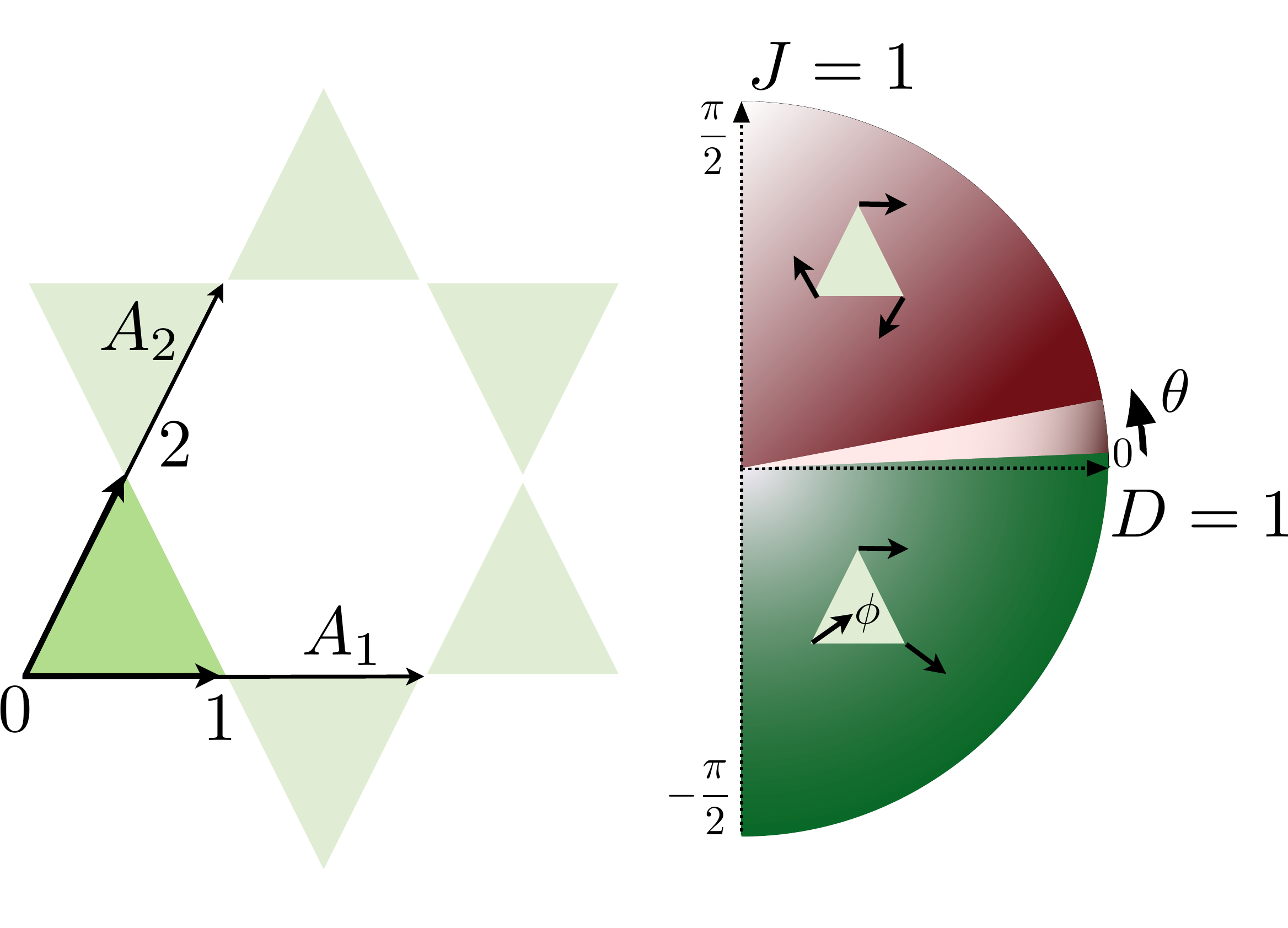}
\caption
{Kagome lattice (left) and the ground-state phase diagram of the model consisting of nearest-neighbour exchange $J=\sin\theta$ and dipolar interactions of strength $D=\cos\theta$ (right). $A_{1},\;A_{2}$ are the basis vectors and a dark green triangle denotes the unit cell of three sites. 
The system exhibits long-range $120^o$ order for $\pi/2\geq\theta>\theta_{1}$ with $ \theta_{1}=10.01^0$ and ferrimagnetic order  for $-\pi/2<\theta<\theta_{2}$ where $\theta_{2}=1.03^{o}$. The latter has two spins inclined with respect to one of the unit-cell edges by angle $\pm\phi(\theta)$. The area between two phases  possibly contains an incommensurate intermediate regime. }
\label{lattice}
\end{figure}

The kagome lattice given in Fig. \ref{lattice} is an Archimedean lattice \cite{richter2004quantum}, a triangular lattice of triangles. The positions of the triangular Bravais lattice points are denoted by $\mathbf{R}^{l}$ while each site in the unit-cell is labeled by $\mathbf{r}_{i}$, so that a site is labeled by $\mathbf{R}^{l}_{j}=(\mathbf{R}^{l}+\mathbf{r}_{j})$. Throughout the paper  the lattice constant $R_{nn}$ is set to $1/2$ such that the full translation of the three site unit cell is the unit of length.

The general Hamiltonian of the system of $N$ spins is 
\begin{eqnarray}
H & = & \sum_{k,i,l,j}\sum_{\alpha,\beta}J_{ij}^{\alpha\beta}(\mathbf{R}_{ij}^{kl})S_{i}^{\alpha}(\mathbf{R}^{k})S_{j}^{\beta}(\mathbf{R}^{l}),\\
\label{hamiltonian}
J_{ij}^{\alpha\beta}\left({\bf {R}}\right)&=&\frac{1}{2}\!\left[\!\!J \delta_{\alpha\beta}+ DR^3_{nn}\!\!\left(\frac{\delta_{\alpha\beta}}{|{\bf{R}}|^{3}}-3\frac{R^{\alpha}R^{\beta}}{|{\bf{R}}|^{5}}\right)\!\!\right].
\label{int_matrix}
\end{eqnarray}
Here $\mathbf{R}_{ij}^{kl}$ is the vector between two interacting classical $O(3)$ spins $S_{i}^{\alpha}(\mathbf{R}^{k})$ and $S_{j}^{\beta}(\mathbf{R}^{l})$, of unit length.  $k$ and $l$ index the unit cell, while $i$ and $j$ run over the sites of the basis in the
unit cell and Greek $\alpha$ and $\beta$ denote the components of the vectors $x,\; y$
and $z$. The first term of the interaction matrix (\ref{int_matrix}) is the nearest-neighbor exchange while the second is the dipole interaction, with $R^3_{nn}$,  the nearest-neighbor distance, included  for normalization. A factor $\frac{1}{2}$ has been included to avoid double counting. $J>0$ is the energy scale of the antiferromagnetic (AFM) nearest-neighbor exchange. The dipolar energy scale is 
\begin{equation}
D=\frac{\mu_{0}}{4\pi}\frac{\mu^2}{R^3_{nn}},
\label{dipolar_scale}
\end{equation}
where $\mu$ is the magnetic moment of the ions. 

We parametrize the relative strength of $J$ and $D$ via an angle $\theta$ (Fig.~\ref{lattice}): 
\begin{equation}
J=\sin{\theta},
D=\cos{\theta},
\label{JDparametrization}
\end{equation}
with the unit of energy set to $J^2+D^2=1$.

Fourier transformation of the Hamiltonian (\ref{hamiltonian})  yields
\begin{eqnarray}
H & = &\sum_{\mathbf{k},i,j}J_{ij}^{\alpha\beta}(\mathbf{k})S_{i}^{\alpha}(-\mathbf{k})S_{j}^{\beta}(\mathbf{k})
\label{hamiltonian_reciprocal}\\
J_{ij}^{\alpha\beta}\left({\bf {k}}\right)&=&\sum_{kl}J_{ij}^{\alpha\beta}(\mathbf{R}_{ij}^{kl})\exp[-i\mathbf{k}\mathbf{R}_{ij}^{kl}].
\end{eqnarray}
We generate the interaction matrix for the dipolar interactions  using Ewald summation \cite{enjalran_gingras2004ewald_pyrochlore}, which we confirm by the direct lattice summation possible in two dimensions.

\section{Luttinger-Tisza analysis}
\label{sec:luttinger_tisza}
\begin{figure}
\includegraphics[trim=2.2cm 2.8cm 2.7cm 2.8cm, clip=true, width=0.235\textwidth,scale=1.]{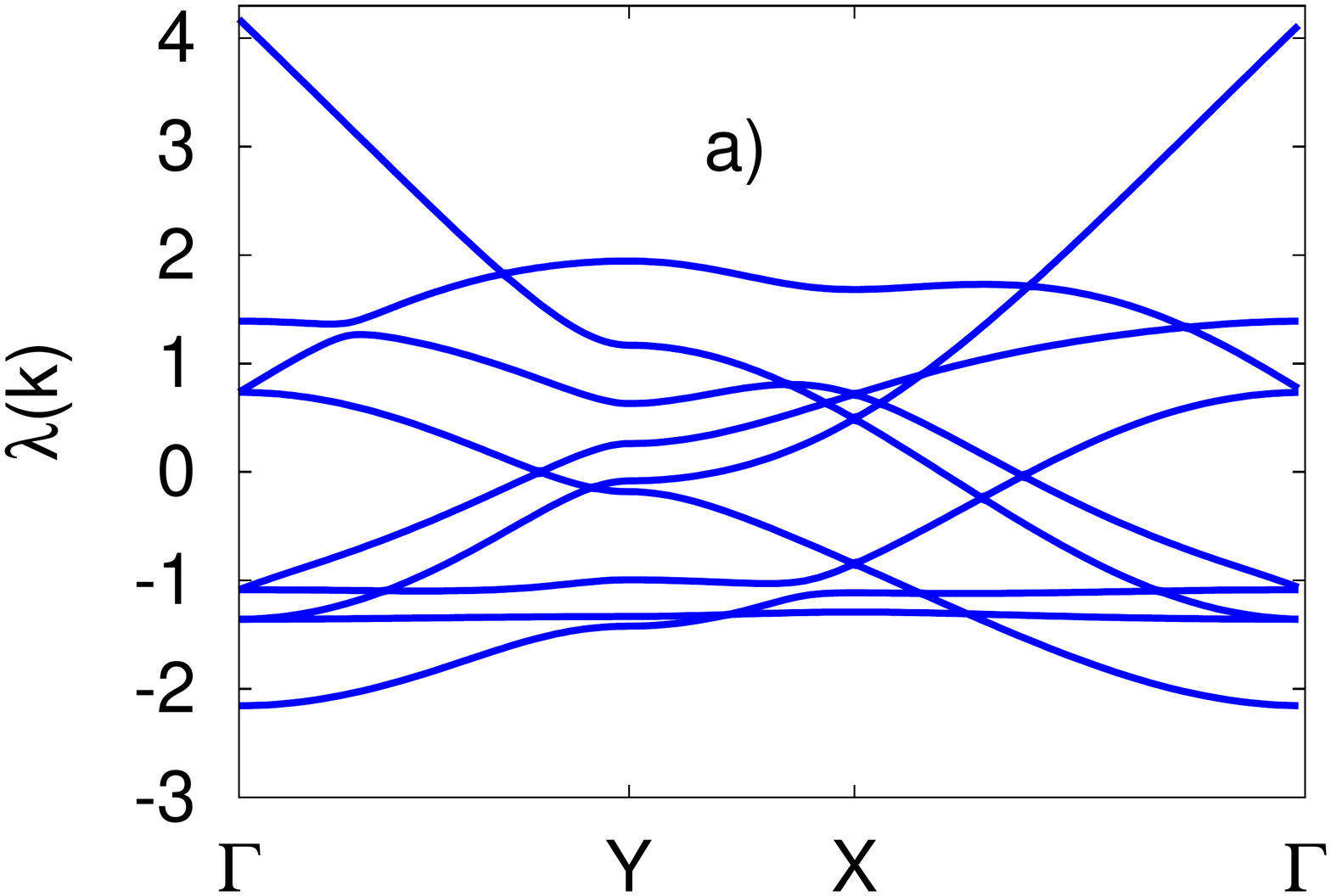}
\includegraphics[trim=2.2cm 2.8cm 2.7cm 2.8cm, clip=true, width=0.235\textwidth,scale=1.]{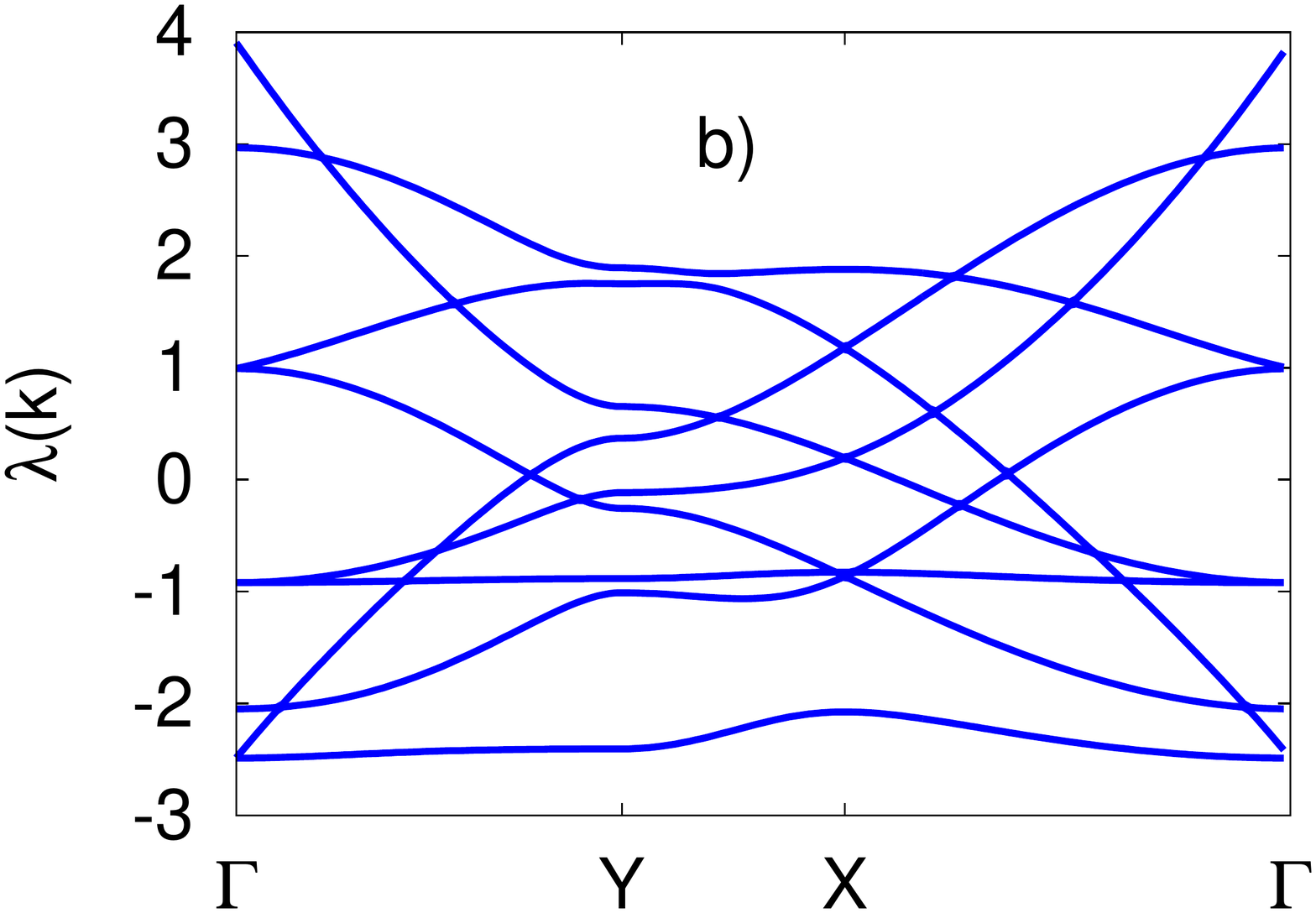}
\caption{Eigenvalues of the interaction matrix $J_{ij}^{\alpha\beta}\!\left({\bf {k}}\right)$ along  lines in the Brillouin zone. a) Spectra for dominant exchange case, $\theta=\pi/4$. b) spectra for $D=1$, $\theta=0$. The eigenvalue lowest in energy is generally at the $\Gamma$ point ${\bf {k}}_0$.}
\label{interaction_matrix_spectrum}
\end{figure}
We first determine a ground state using the Luttinger-Tisza (LT) method \cite{luttinger_tisza} where it applies. 
Decomposing
the interaction matrix into its Fourier components, and   denoting by $\lambda_{min}(\mathbf{k})$ the lowest eigenvalue(s) of the
interaction matrix, we use the fact that the energy of {\em any} spin configuration satisfies the bound
\begin{equation}
H \geq N \lambda_{min}(\mathbf{k}_{0}). 
\end{equation} 
If there exists a spin configuration which can be decomposed into a linear combination of 
{\em only} the 'optimal' \cite{lapa_henley2012}
LT eigenvectors corresponding to these eigenvalues, it is a global ground state. This happens if the 
 ``strong constraint'' of unit length for the spins
 \begin{equation}
|\textbf{S}_{i}|^2=1
\label{strong_constraint}
\end{equation}
does not conflict with the optimal eigenvectors, which however in general have entries 
with different amplitudes. 
In the latter case, not unusual for non-Bravais lattices, 
non-optimal modes have to be admixed, and the LT approach only yields an (often
rather useful) guess at possible ground states, or at least at leading instabilities from the high-temperature paramagnet.

\subsection{Dominant nearest-neighbor exchange.}
For pure nearest-neighbor antiferromagnetic exchange $\theta=\pi/2$ ($D=0$), the lowest branch of the interaction matrix is  exactly flat (dispersionless) reflecting the high  ground state degeneracy\cite{chalker1992kagome,harris1992_kagome, ritchey1993spin,huse1992classical}. Decreasing $\theta$ we move to nonzero $D>0$ which immediately lifts the degeneracy, selecting a ground state at wavevector ${\bf{k}}_{0}=(0,0)$, as shown in Fig. \ref{interaction_matrix_spectrum} a) for $\theta = \pi/4 $. The optimal eigenvector satisfies the constraint (\ref{strong_constraint}) and results in a $120^o$ state which is doubly degenerate reflecting two possible chiralities. Further increase of $D$ leads to level crossing at $\theta_{1} = 10.01^{0}$. Hence the $120^o$ state is certainly stable up to this point, as we have also confirmed in Monte-Carlo simulations. 
\subsection{Dominant dipolar exchange.}
For $\theta<\theta_{1}$ LT no longer yields an  exact ground state \footnote{In this region we also observe a shift of optimal wave vector ${\bf{k}}_{0}$ to the $Y$-point if $\theta_{1}<\theta<\theta_{Y}$, where $\theta_{Y} =9.77^{0}$.}. Instead,  we enter an intermediate regime where neither spin-wave nor Monte-Carlo computations (see Sections \ref{sec:spin_waves} and \ref{section_mc_simulation}) allow us reliably to conclude on the ground state. This regime persists up to the point $\theta_{2}=1.03^{0}$ beyond which the $120^o$ state is no longer even a stable local minimum at  ${\bf{k}_0}=(0,0)$. 

For purely dipolar interactions $\theta=0$ the minimal eigenvalue $\epsilon_{0} = -2.487$ is doubly degenerate and again occurs at $\textbf{k}_{0}=(0,0)$, Fig. \ref{interaction_matrix_spectrum} b.   The best state we find has  two of the spins are inclined approximately by $\phi_\approx \pm 16^0$ with respect to one of the unit-cell edges while the third spin remains unchanged (right panel of Fig. \ref{lattice}).  This situation persists, with varying 
$\phi(\theta)$ until the ferromagnetic point $\theta=-\pi/2$.
However, in general
no combination of the pair of eigenvectors satisfies the strong constraint on spin length (\ref{strong_constraint}). To determine the true ground state for hard unit length spins, we thus need to allow the admixing of other modes, so that we next turn to numerics. 

\section{Numerical energy minimization for $D \gg J$}.
\label{sec:numerical_minimization}
\begin{figure}
\centering
\includegraphics[trim=2.2cm 2.3cm 1.9cm 2.8cm, clip=true, width=0.7\columnwidth]{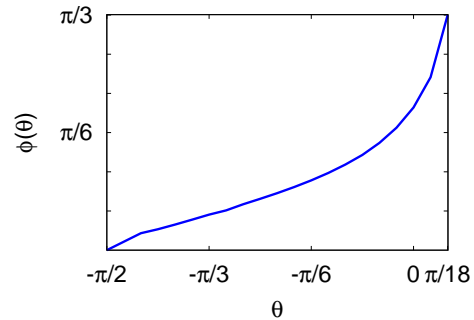}\caption
{Inclination angle $\pm\phi$ of two of the spins in the unit cell (Fig. \ref{lattice}) as function of $\theta$ (Eq. \ref{JDparametrization}) in the ferrimagnetic phase.}
\label{fig:phi_vs_theta}
\end{figure}
Our Monte-Carlo simulations (Section \ref{section_mc_simulation}) do unveil an $\textbf{k}_{0}=(0,0)$ ordering at low temperatures,  suggesting that hard spin constraint may optimally be satisfied by admixing higher modes at $\textbf{k}_{0}=(0,0)$ only. We therefore constrain our problem to a single unit cell and perform a numerical minimization of the Hamiltonian (\ref{hamiltonian_reciprocal}). 
The minimal energy configuration for the single unit cell is indeed the state found in full lattice Monte Carlo simulations.
The ground state is a ferrimagnetic configuration in which the spins ${\bf{S}}_{i}$ take the following angles with one of the three edges in the unit cell
\begin{align*}
\phi_{1} & =\phi,\;\phi_{2}=-\phi,\;\phi_{3}=0,\\
\end{align*}
with
\begin{equation}
\phi\approx36.42^{0}
\end{equation}
As we change $-\pi/2<\theta<\theta_{2}$ we can obtain a minimal energy ferrimagnetic configuration with a drifting $\phi(\theta)$, as shown in Fig. \ref{fig:phi_vs_theta}.

\section{Linear spin-wave theory}
\label{sec:spin_waves}
\begin{figure}
\includegraphics[trim=0.7cm 2.1cm 0.8cm 0.cm, clip=true,width=0.23\textwidth,scale=0.9]{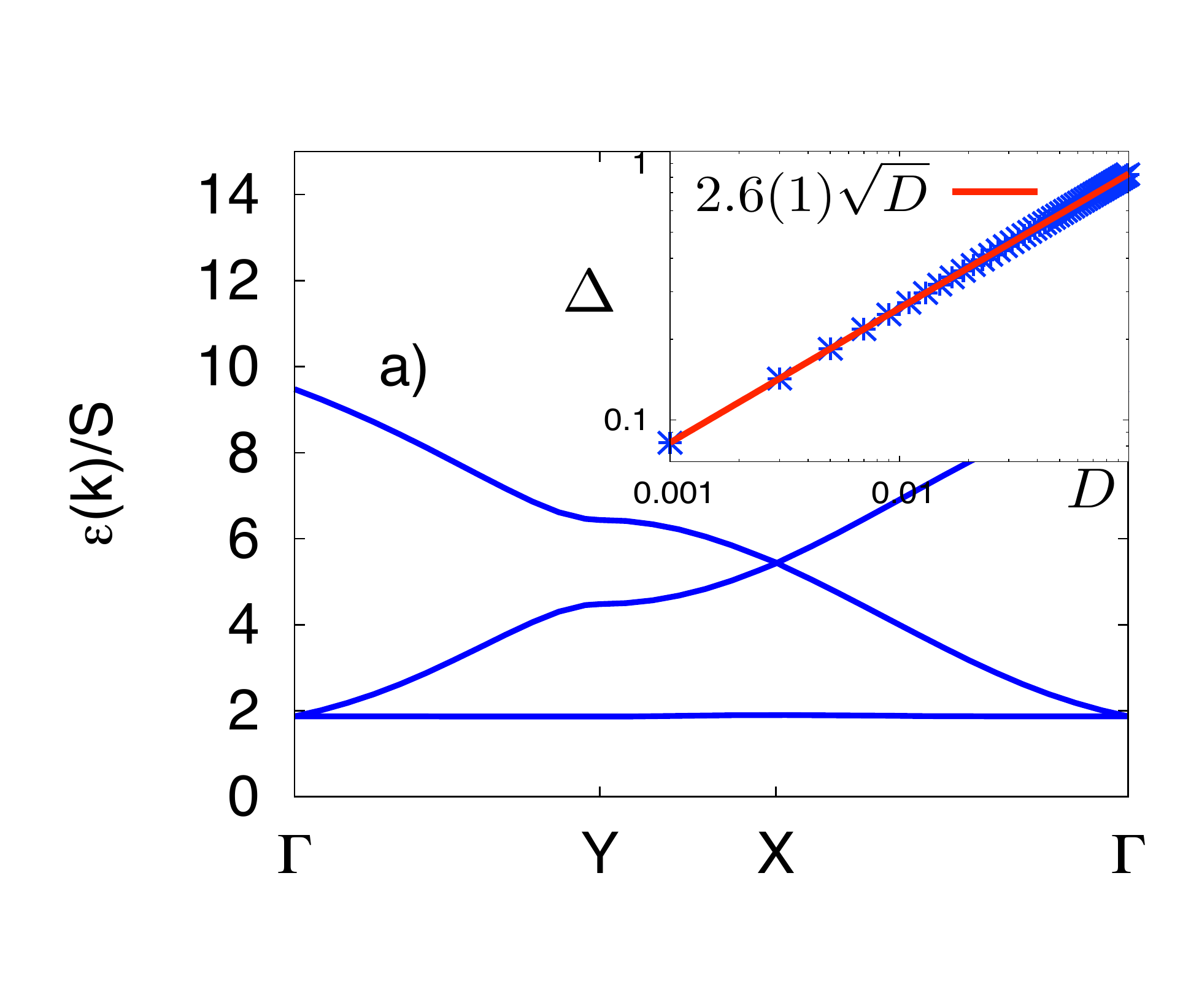}
\includegraphics[trim=2.2cm 2.8cm 2.4cm 2.85cm, clip=true,width=0.23\textwidth,scale=0.9]{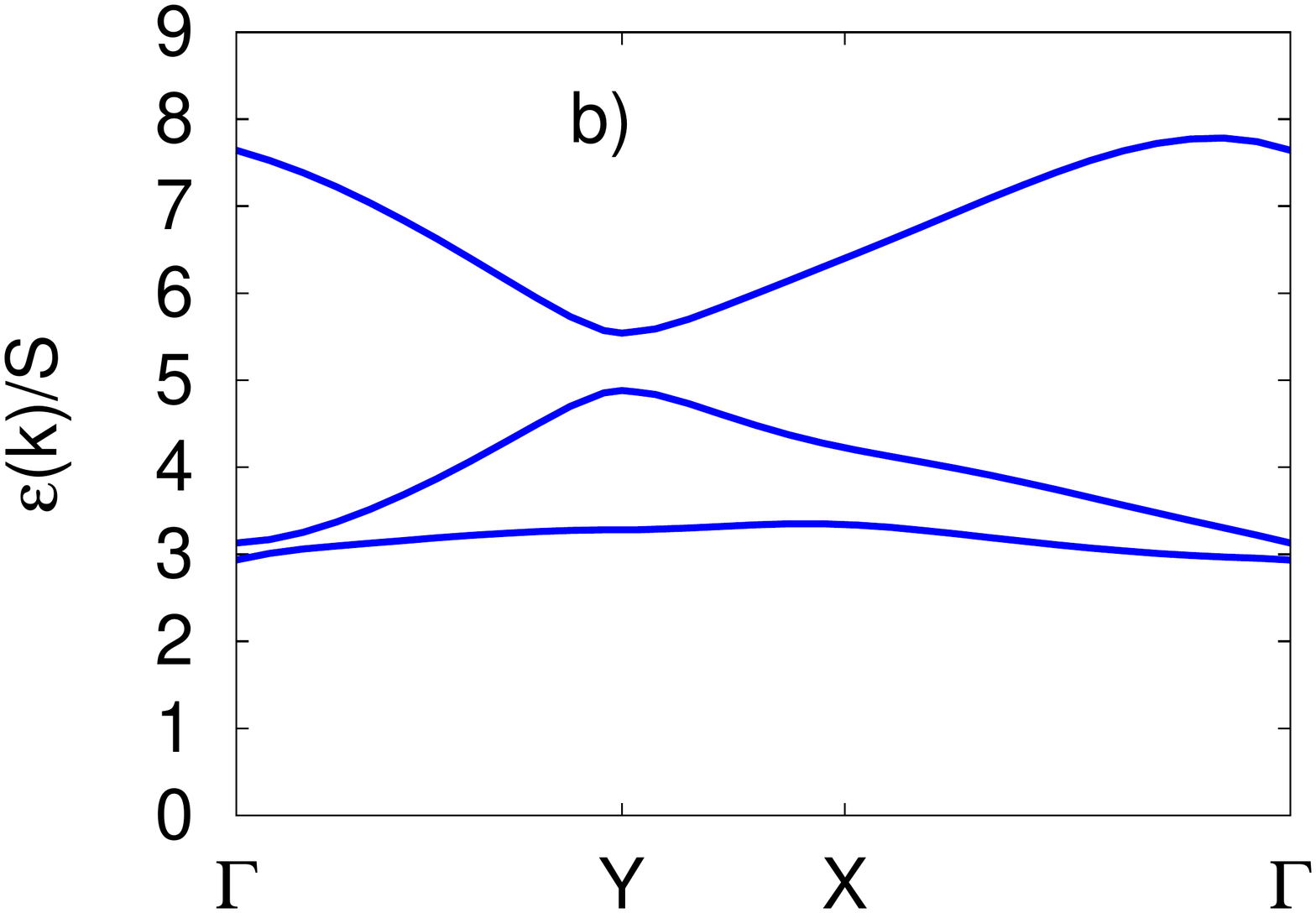}
\caption{Spin-wave spectra for a) $120^0$ state at $\theta=\pi/4$  and for b) the ferrimagnetic ground state obtained by energy minimization at $\theta=0$. In a), the lowest
branch remains almost perfectly flat, while acquiring a gap $\propto\sqrt{D}$ (inset).}
\label{fig:sw_spectrum}
\end{figure}
\begin{figure}
\includegraphics[trim=2.cm 2.1cm 2.4cm 2.7cm, clip=true,width=0.23\textwidth,scale=0.9]{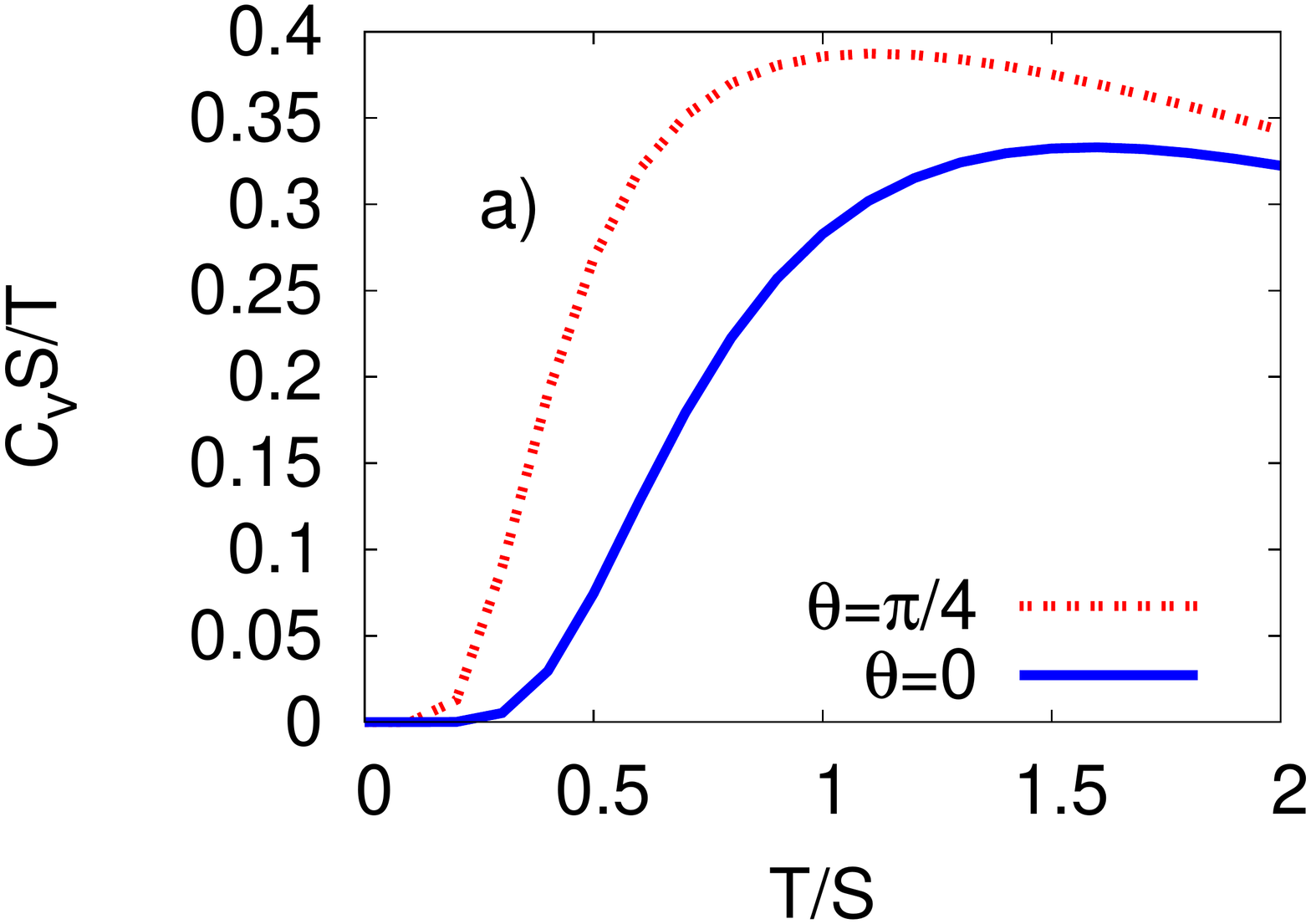}
\includegraphics[trim=2.cm 2.1cm 2.4cm 2.7cm, clip=true,width=0.23\textwidth,scale=0.9]{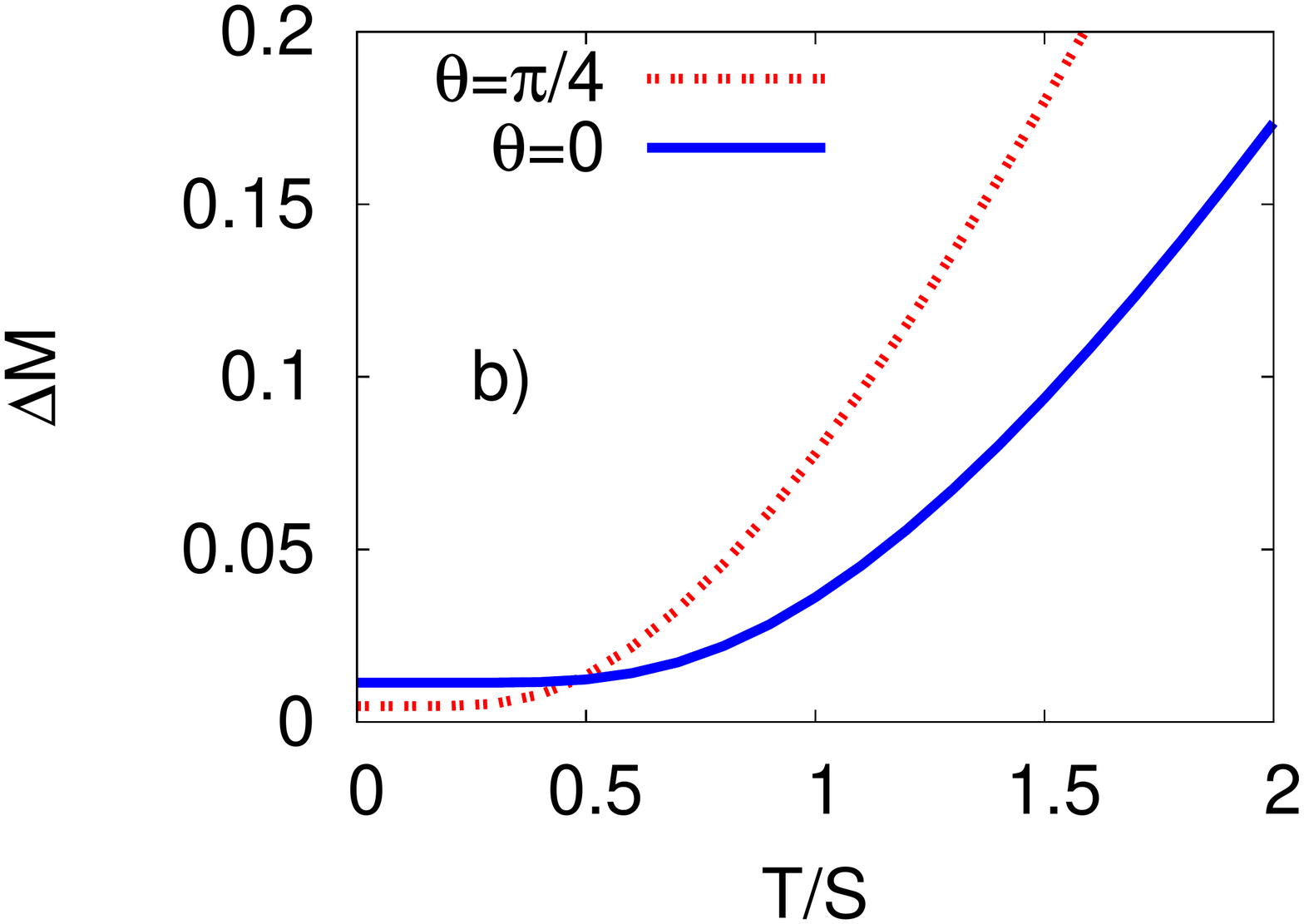}
\caption{a) Specific heat scaled by the temperature and b) temperature dependence of the reduction of sublattice magnetization due to quantum fluctuations calculated using linear spin wave theory,
which is controlled in the limit of small $T$.}
\label{fig:sw_thermodynamics}
\end{figure}
We next study the role of quantum fluctuations around the two ground states discussed above. 
We find that both states are locally stable, and exhibit a lowest band with little dispersion, in particular
for the 120$^o$ state. 

We evaluate the spin-wave spectrum of non-collinear spin structures using standard methods \cite{white1965diagonalization, sw_maestro_gingras, colpa1978diagonalization}. The Hamiltonian of a Bose gas of magnons 
reads
\begin{eqnarray}
H&=&H^{(0)}+\sum_{\mathbf{k}}\sum_{i}\epsilon_{i}(\mathbf{k})\\
&&+\sum_{\mathbf{k}}\sum_{i}\epsilon_{i}(\mathbf{k})\left[a_{i}^{\dagger}(\mathbf{k})a_{i}(\mathbf{k})+a_{i}^{\dagger}(-\mathbf{k})a_{i}(-\mathbf{k})\right]\nonumber,
\label{spin_waves_ham}
\end{eqnarray}
where $a_{i}(\mathbf{k})$ are boson annihilation operators with $H^{(0)}$ the classical ground state energy. For a stable ground state spin configuration, $H$ is Hermitian and all the spin-wave  eigenenergies $\epsilon_{i}(\mathbf{k})$ are real. 
This yields the specific heat  $C_{v}(T)$ and  magnetization $M(T)$, allowing in principle for comparison with experimental data at low temperature, e.g. below a scale set by the gap in the excitation spectrum \cite{quilliam2007_sw_therm_pyrochlore}.

We first confirm that the $120$-degree and ferrimagnetic states at $\theta=\pi/4,0$, respectively,  are stable to quantum fluctuations. 
While it is known from previous studies \cite{harris1992_kagome} that for the $120^o$ state the spin-wave excitation spectrum has a fully dispersionless (flat) band at zero energy as well as two-fold degenerate  
acoustic mode, the addition of $D$  leads to a gap in the excitation spectrum 
proportional to $\sqrt{D}$ at small $D$.   We plot the spin-wave spectra for cases $\theta=\pi/4$ and $\theta=0$ in Fig. (\ref{fig:sw_spectrum}) where it is clearly seen that in both cases the leading effect of dipolar interactions is pushing the zero modes to finite frequency, expected on account of the absence of a continuous symmetry. The dispersion  of the lowest branch of the $120^0$ state is only weakly affected. This fact can manifest itself in finite energy almost ${\bf{k}}$-independent resonance in inelastic neutron scattering \cite{sw_maestro_gingras,lee_broholm_kim2000}. The existence of the gap in the spectrum affects the corresponding specific heat and sublattice magnetization (Fig.~\ref{fig:sw_thermodynamics}). The gap leads to an exponential suppression of specific heat $C_{v}\sim\exp[-\Delta/T]$ or reduction of staggered magnetization $\Delta M(T)\sim \exp[-\Delta/T]$ with the gap $\Delta$.

Moreover we have checked for both $\theta = 0, \pi/4$ that these are the only stable spin-configurations at $\textbf{k}_{0}=(0,0)$. We close this section by noting that  both the $120^0$ and ferrimagnetic states are locally stable within the boundaries of intermediate phase (Fig. \ref{lattice}). 

\section{Monte-Carlo simulations}
\label{section_mc_simulation}
This section pursues two goals. Firstly, the  ground states are confirmed numerically; secondly, 
the corresponding finite-temperature phase transitions are analysed in detail. This is done with 
computationally intensive but tractable Monte-Carlo simulations of the system on 
finite lattices with linear dimension  of $L\leq24$ unit cells or $N\leq1728$ sites.

\begin{figure}
\includegraphics[page=1,trim=2.cm 2.3cm 2.7cm 2.8cm, clip=true,width=0.23\textwidth]{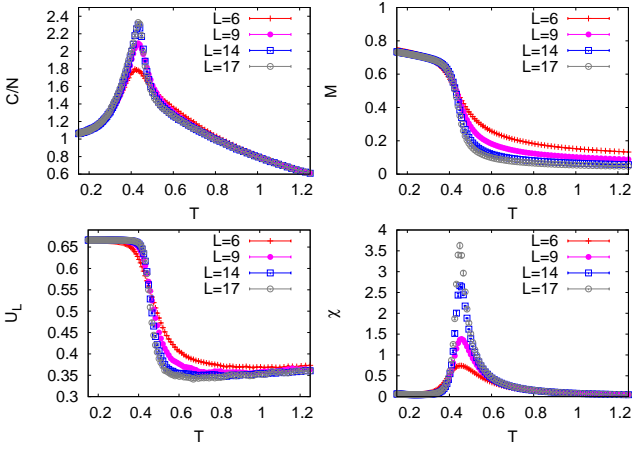}
\includegraphics[page=2,trim=2.cm 2.3cm 2.7cm 2.8cm, clip=true,width=0.23\textwidth]{mc_thermodynamics.pdf}
\includegraphics[page=3,trim=2.cm 2.3cm 2.7cm 2.8cm, clip=true,width=0.23\textwidth]{mc_thermodynamics.pdf}
\includegraphics[page=4,trim=2.cm 2.3cm 2.7cm 2.8cm, clip=true,width=0.23\textwidth]{mc_thermodynamics.pdf}
\caption{Monte-Carlo results for specific heat $C/N$, Binder cumulant $U_{L}$, magnetization $M$ and susceptibility $\chi$ for $\theta=0$ ($D=1$ and $J=0$). }
\label{mc_thermodynamics_ferri}
\end{figure}
\begin{figure}
\includegraphics[page=5,trim=2.cm 2.3cm 2.7cm 2.8cm, clip=true,width=0.23\textwidth]{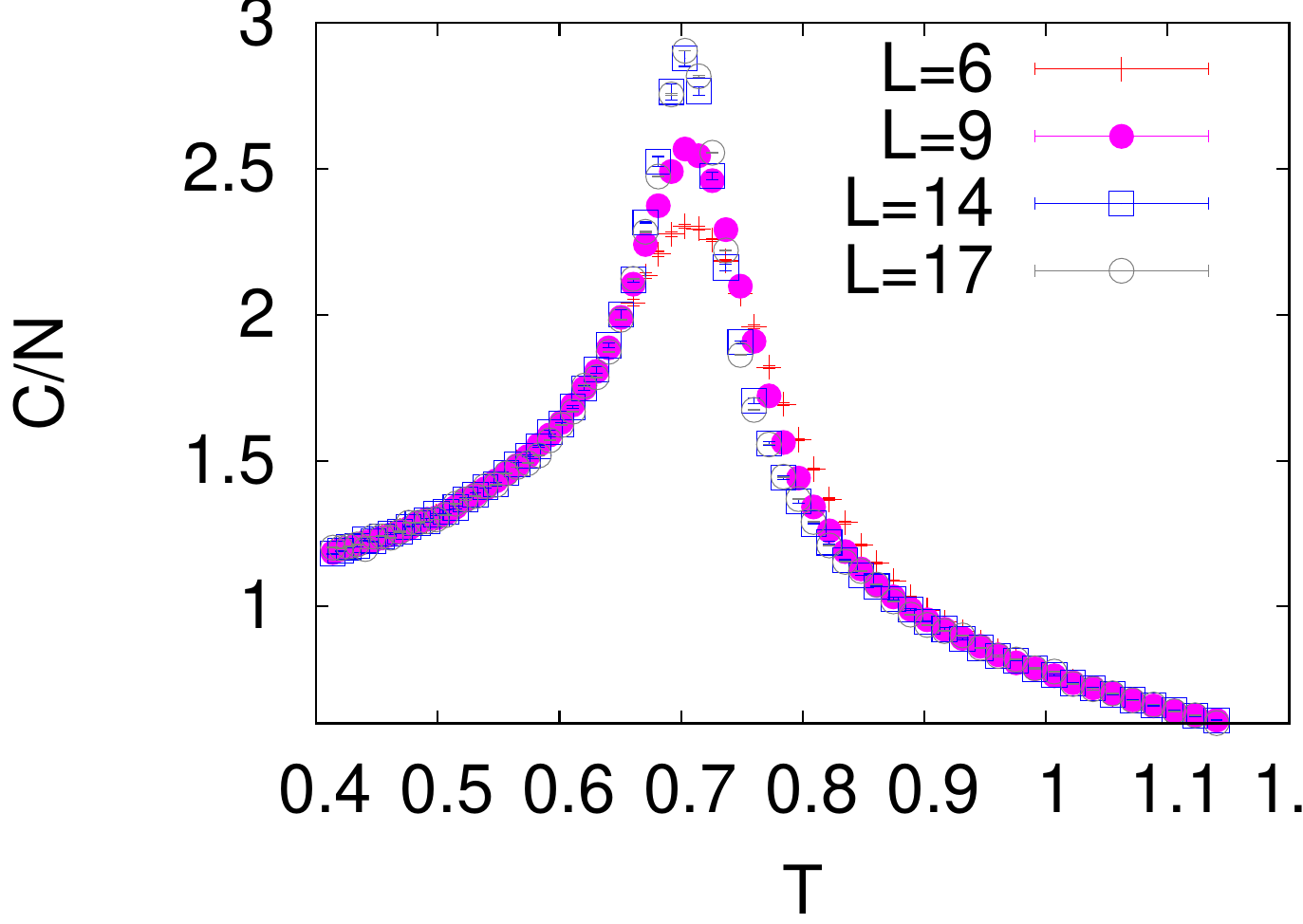}
\includegraphics[page=6,trim=2.cm 2.3cm 2.7cm 2.8cm, clip=true,width=0.23\textwidth]{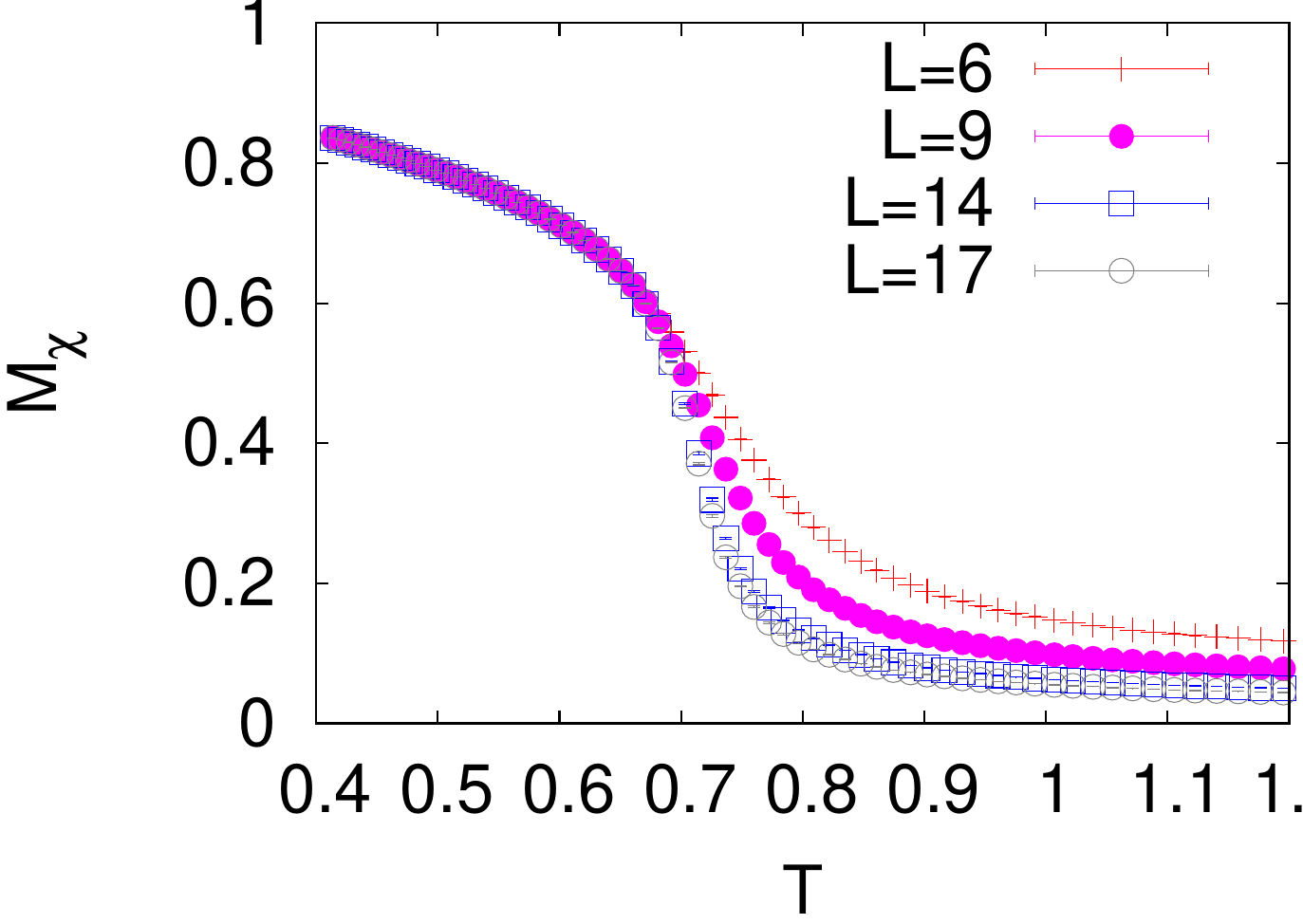}
\includegraphics[page=7,trim=2.cm 2.3cm 2.7cm 2.8cm, clip=true,width=0.23\textwidth]{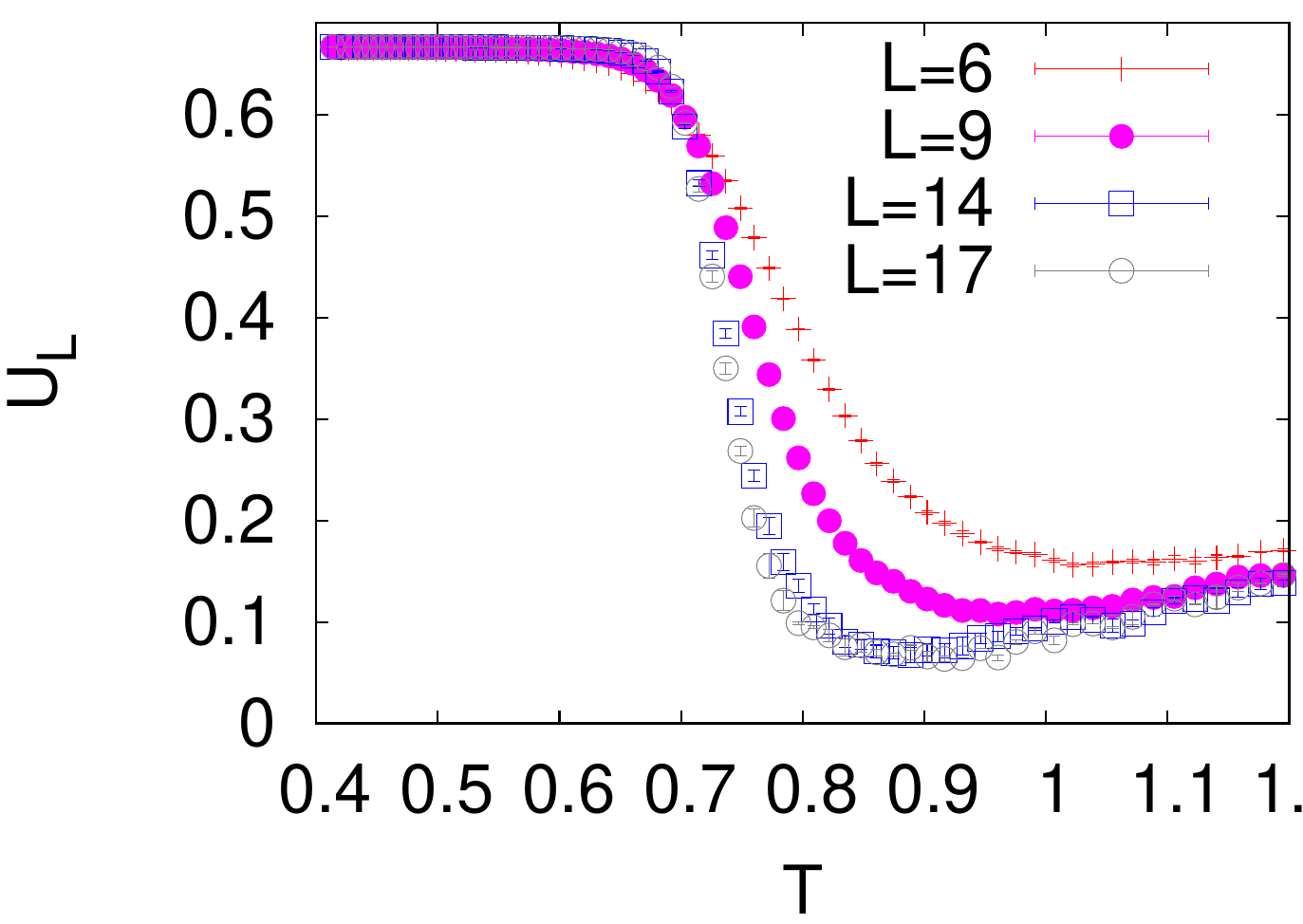}
\includegraphics[page=8,trim=2.cm 2.3cm 2.7cm 2.8cm, clip=true,width=0.23\textwidth]{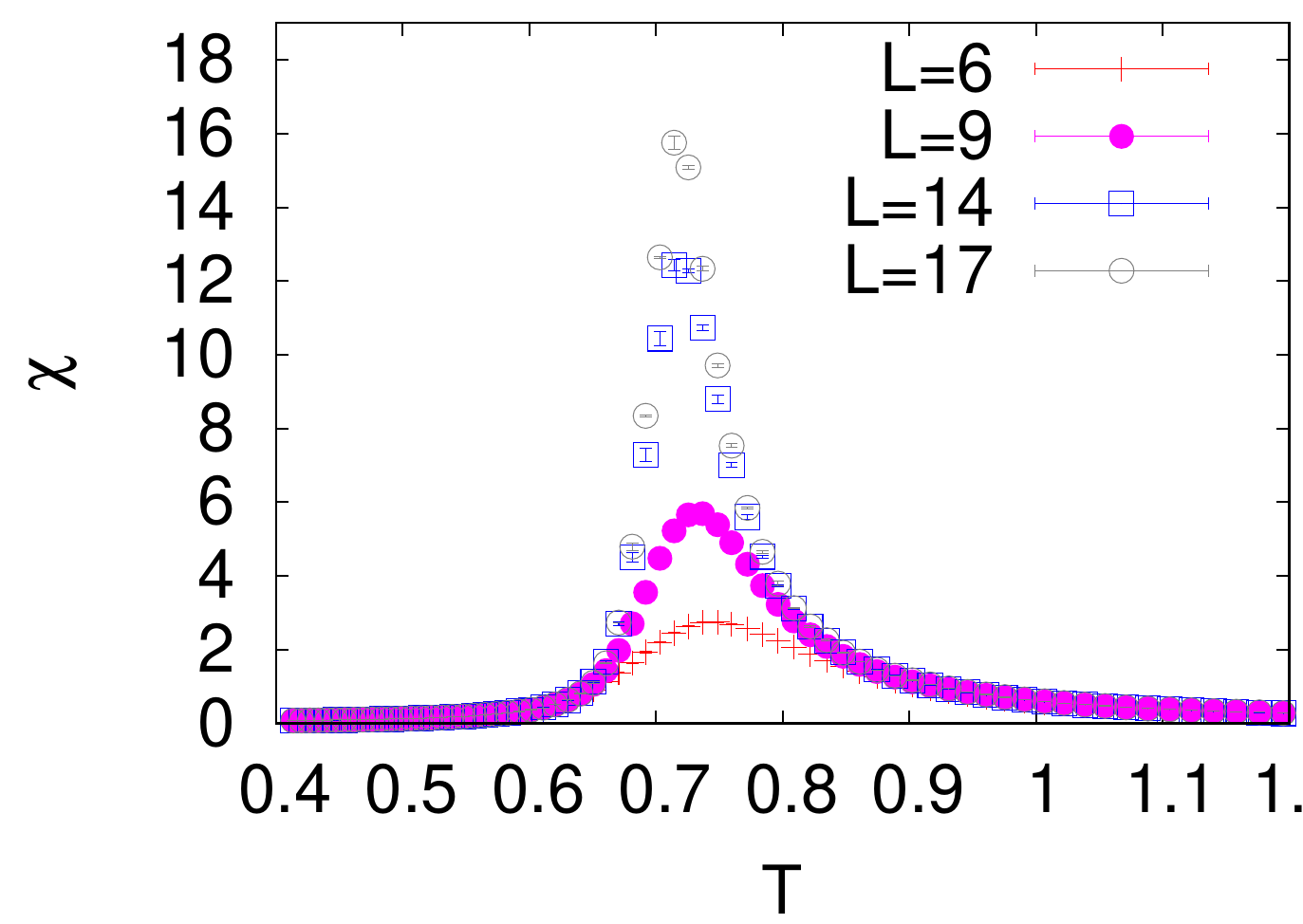}
\caption{Monte-Carlo results for specific heat $C/N$, Binder cumulant $U_{L}$, chiral order parameter $M_{\chi}$ and corresponding susceptibility $\chi$ for $\theta=35.6^{0}$. }
\label{mc_thermodynamics_chiral}
\end{figure}
%
We employ parallel tempering with $64$ to $128$ replicas in the temperature range $T=0.125-2.95$ for the phase transition analysis and in the range $T=0.00625-2.95$
to investigate the low energy configuration of the dipoles.  
One Monte-Carlo step
corresponds to a sweep over the lattice in which on average every
spin is touched. We perform $\approx10^{6}$ Monte Carlo steps for the thermalization, 
 followed by $\approx10^{6}$ steps for every measuring round.

We obtain thermodynamic properties of the model (specific heat, uniform and staggered -- $120^0$ state --magnetization, magnetic susceptibility, fourth order Binder cumulant) as well as the  structure of the low-temperature spin configuration.  

For the set of parameters leading to the ferrimagnetic ground state we analyze the phase transition via the behavior of 
the magnetic order parameter 
\begin{equation}
{\bf{M}}=\frac{1}{N}\sum_{i}(S^{x}_{i},S^{y}_{i},S^{z}_{i}),
\label{ferro_order}
\end{equation}
where the sum is taken over all the sites in the lattice. For the planar $120^0$ ground state order we investigate the order parameter which captures the particular chiral spin pattern,
\begin{equation}
{\bf{M}_{\chi}}=\sqrt{{\bf{m}}_{\chi}{\bf{m}}_{\chi}^{*}},
\label{ferro_order}
\end{equation}
where
\begin{equation}
{\bf{m}}_{\chi}=\frac{1}{N}\sum_{\mathbf{R}+\mathbf{r}_{j}}{\bf{S}}(\mathbf{R}+\mathbf{r}_{j})\exp({i\phi_{j})},
\end{equation}
and $\phi_{j}$ are sublattice phase angles $\phi_{1}=0$, $\phi_{2}=2\pi/3$, and $\phi_{3}=4\pi/3$.

To investigate the corresponding finite-temperature phase transition we also compute the fourth order Binder cumulant
\begin{equation}
U_{L}=1-\frac{1}{3}\frac{\langle O \rangle^4}{\langle O^2 \rangle^2},
\end{equation}
as well as susceptibility 
\begin{equation}
\chi=\frac{N}{T}\left(\langle O^2 \rangle- \langle O\rangle^2\right),
\end{equation}
and specific heat per spin
\begin{equation}
C/N=\frac{1}{N}\frac{1}{T^{2}}\left(\langle E^2 \rangle- \langle E\rangle^2\right).
\end{equation}

To characterize the phase transitions we employ standard finite-size scaling 
\begin{eqnarray}
\tilde{M}(L^{1/\nu}t)=L^{\beta/\nu}M_{L},\nonumber \\
\tilde{\chi}(L^{1/\nu}t)=L^{-\gamma/\nu}\chi_{L},\nonumber\\
\tilde{C}(L^{1/\nu}t)=L^{-\alpha/\nu}C_{L},
\end{eqnarray}
where $t=(T-T_{c})/T_{c}$ is the reduced temperature. To obtain the critical exponent $1/\nu$ and critical point $T_{c}$ we 
use the scaling relation for the Binder cumulant 
\begin{equation}
\tilde{U}(L^{1/\nu}t)=U_{L}.
\end{equation}
We extract $\nu,\;\beta,\;\alpha,\;\gamma$ and $T_{c}$ via data collapse.

\begin{figure}
\includegraphics[trim=2.cm 2.3cm 2.7cm 2.8cm, clip=true,width=0.23\textwidth]{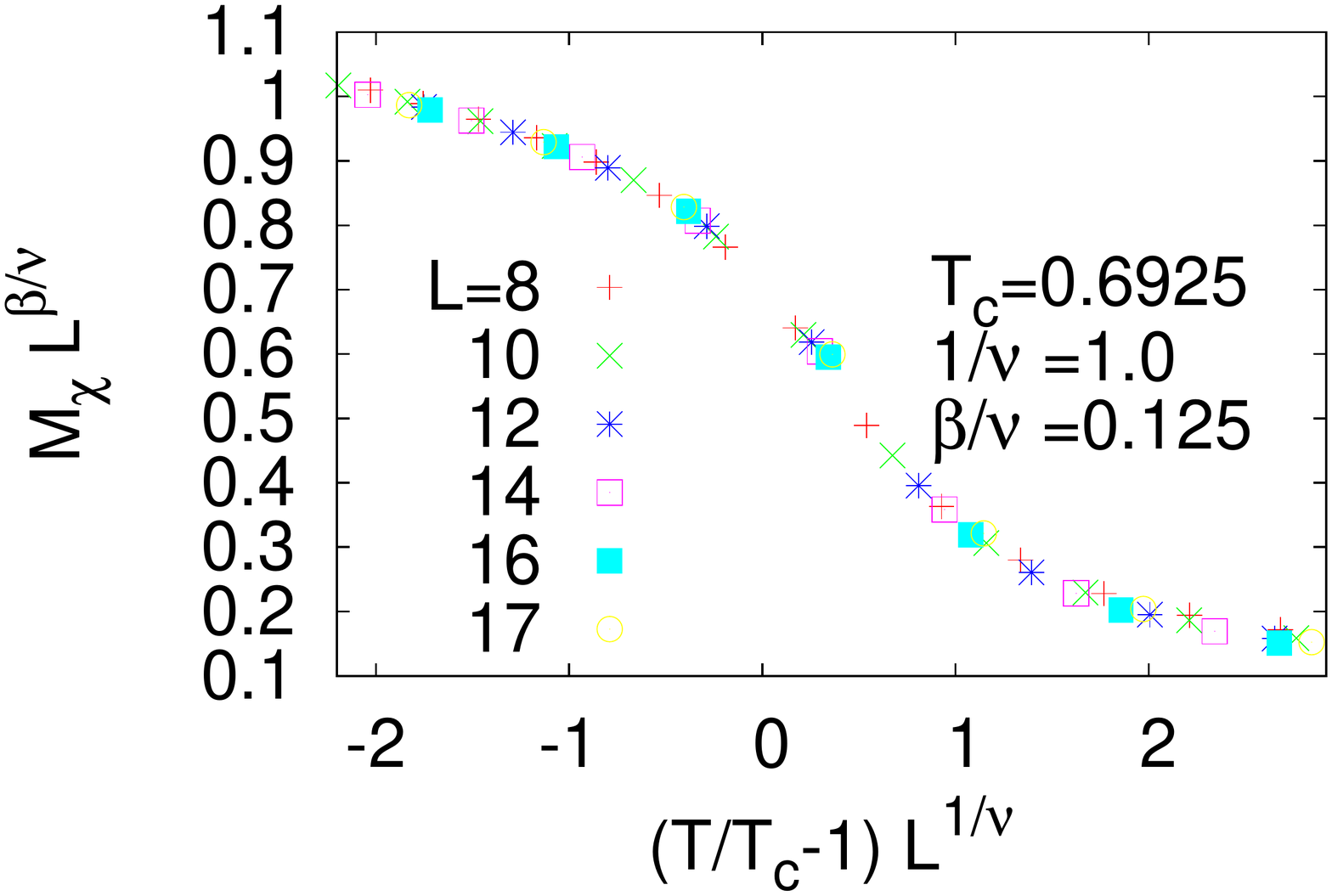}
\includegraphics[trim=2.cm 2.3cm 2.7cm 2.8cm, clip=true,width=0.23\textwidth]{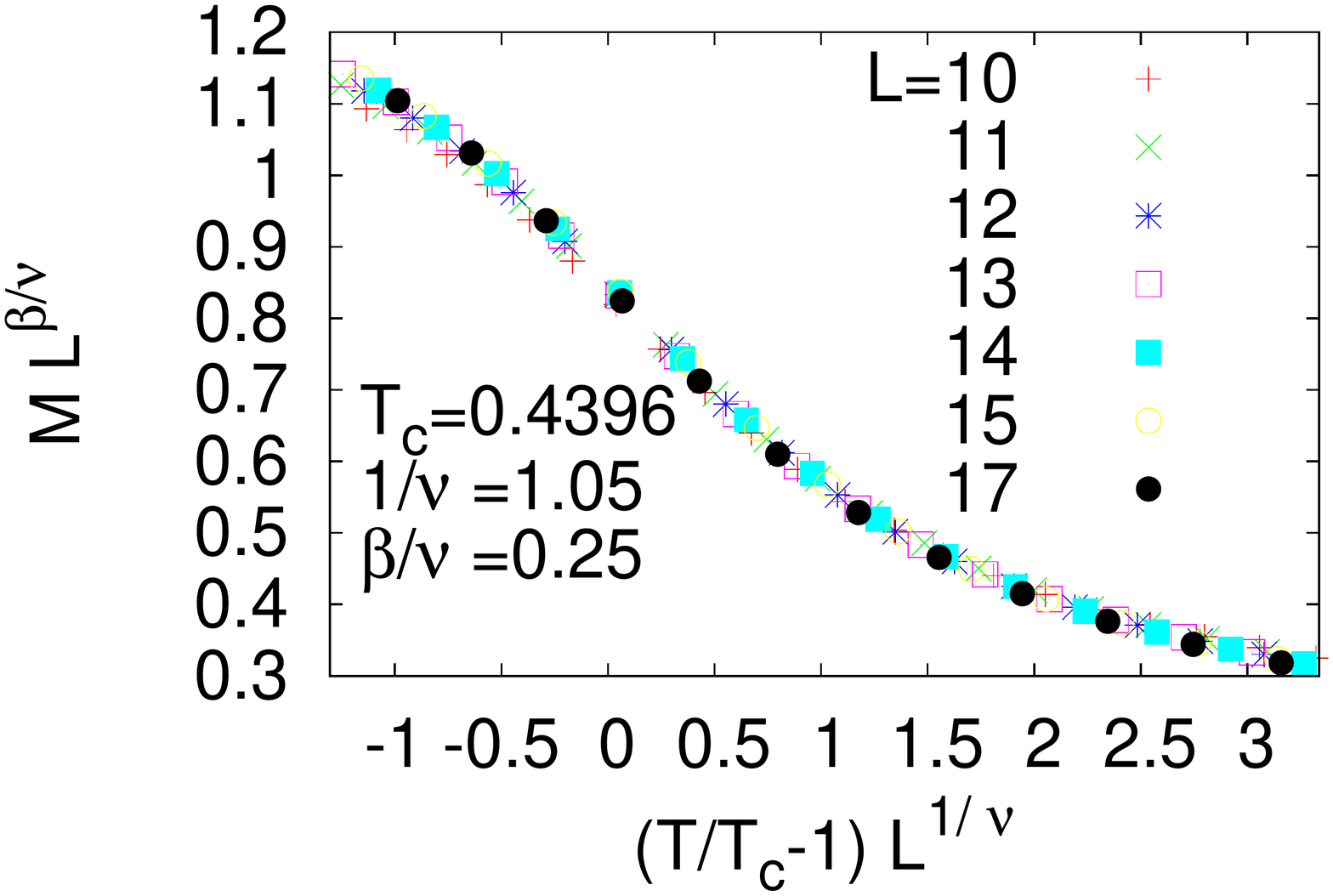}
\includegraphics[trim=2.cm 2.3cm 2.7cm 2.8cm, clip=true,width=0.23\textwidth]{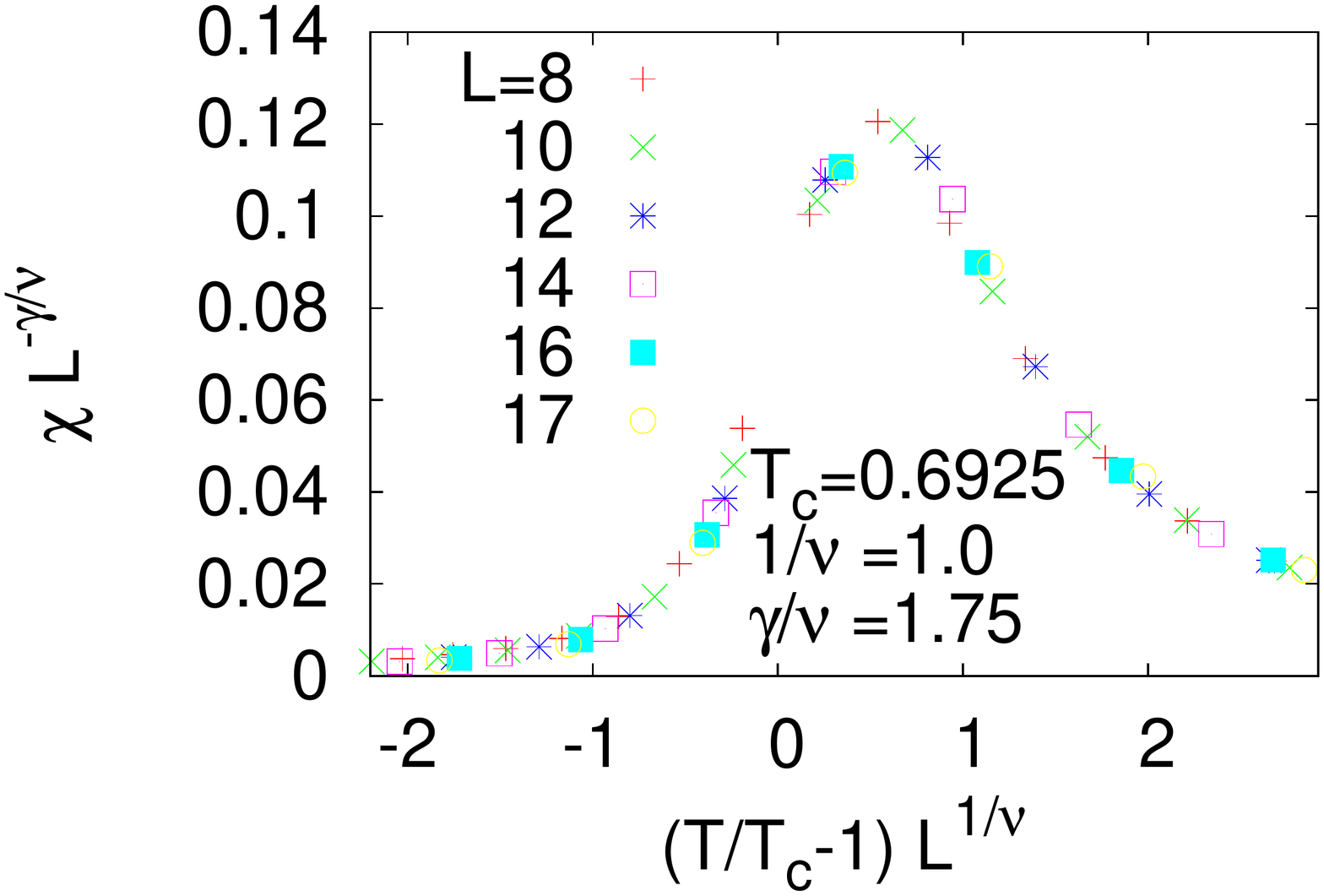}
\includegraphics[trim=2.cm 2.3cm 2.7cm 2.8cm, clip=true,width=0.23\textwidth]{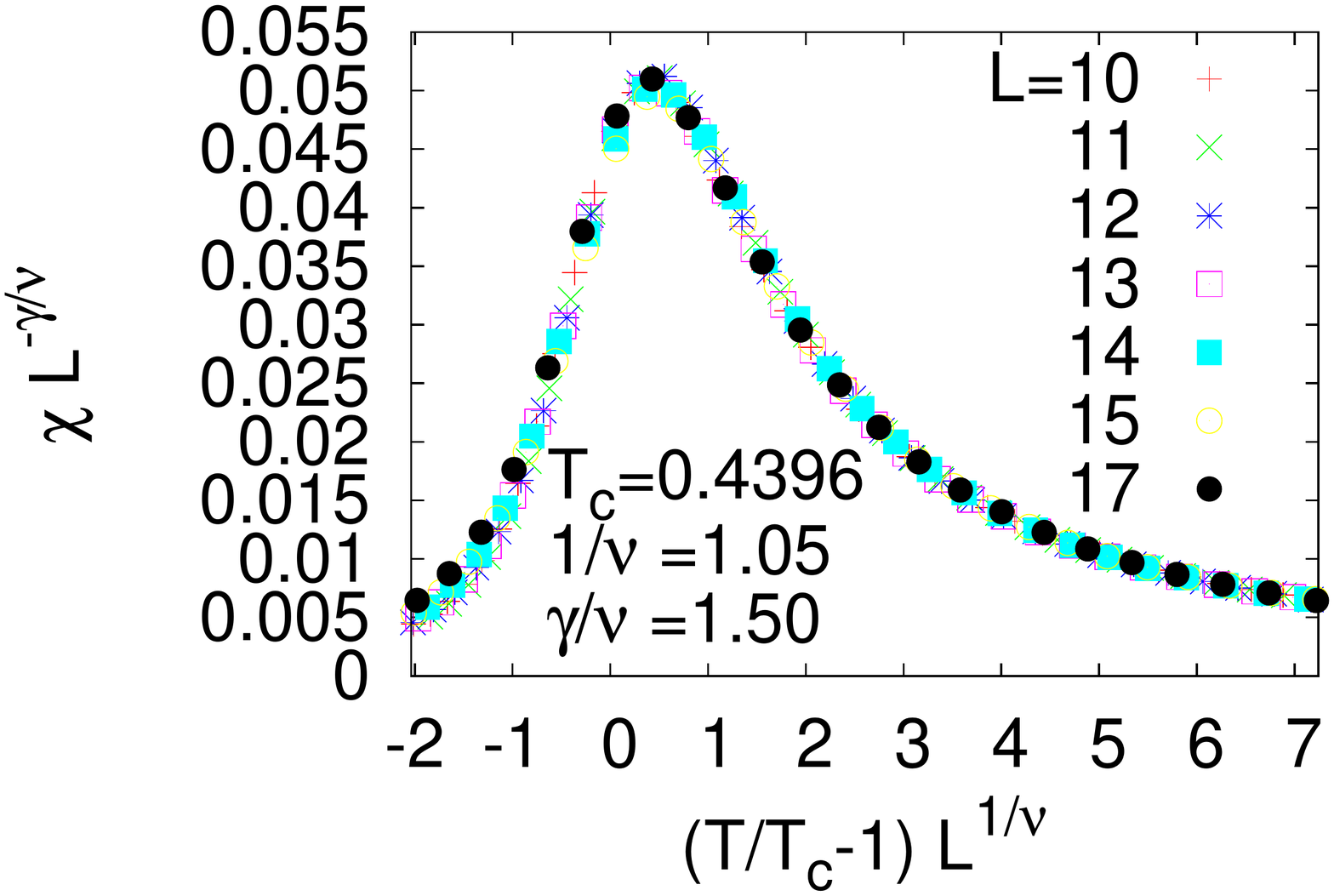}
\caption{Scaling collapse for the magnetic (chiral) order parameter and its susceptibility for the transition 
into the $120^0$ state (left) and the ferrimagnetic state (right).}
\label{fig:data_collapse}
\end{figure}

Let us first look at the finite temperature transition to the $120^0$ state. We perform MC simulations deep in the ordered phase for $\theta=35.6^0$. Collapsing the curves for chiral order parameter, Binder cumulant, susceptibility and specific heat yields critical temperature $T_{c}$ as well as the full set of critical exponents $\nu,\;\beta\;,\gamma,\;\alpha$ see Fig. \ref{fig:data_collapse}. The transition occurs at  
$T_{c}=0.692(5)$, consistent with the  2D Ising 
universality class with critical exponents $\nu=1,\;\beta=1/8,\;\gamma=7/4,\;\alpha=0$, reflecting the discrete $Z_{2}$ symmetry of the chiral order parameter. 

For dominant dipolar interactions we analyze two points, $\theta=0$, and for $\theta=1^0$. The ferrimagnetic order  has a six-fold discrete symmetry and at $T=0$ differs only in the angle $\phi$ of the two inclined spins (Fig. \ref{lattice} b)). We therefore expect the corresponding transitions  to belong to the same universality class. 

Our Monte-Carlo data show a clear divergence of the ferromagnetic order parameter, specific heat and susceptibility as well as crossings of fourth-order Binder cumulant curves. This 
suggests a single second-order phase transition from a high-temperature paramagnet to a low-temperature ferrimagnetic phase (Fig. \ref{mc_thermodynamics_ferri}). Both for $\theta=0$ and $\theta=1^0$  we can extract critical temperatures $T_{c}=0.439(2)$ and $T_{c}=0.406(5)$ as well as the set of exponents which lead to the best data collapse of Binder cumulant, magnetization, susceptibility and specific heat (Table \ref{table_critical_exponents}). Note that correlation length exponent $\nu$ and order parameter exponent $\beta$ increase monotonically with $J$ while the ratios $\beta/\nu\approx0.25$ and $\gamma/\nu\approx1.5$ remain constant with a two-dimensional scaling law implying $\eta=2\beta/\nu\approx0.5$. 

This appears to provide an example of the so-called "weak universality" hypothesis which states that ratios of exponents should be independent of the details of system Hamiltonian with $\eta=2\beta/\nu$ and $\gamma/\nu$  universal while $\alpha$ and $\beta$ are allowed to change \cite{suzuki1974new}. The "weak universality" behavior is often observed as a drift from Berezinskii-Kosterlitz-Thoules (BKT) exponents to discrete (i.e. Ising, Potts) transition exponents \cite{seabra_triangular_heis_field,taroni2008universal}, and may of course
be related to the existence of a large lengthscale. Note that in our case,  we have a correspondence to a six-state clock model arising from a Hamiltonian with both nearest-neighbor and long-range interactions. Individually, a six state clock model with only the former exhibits two KT transitions (not observed here) \cite{clock_model_2D,clock_model_triangular,clock_model_tobochnik,clock_model} while mean-field studies for the case of long range dipolar interactions suggest a single second order low-temperature phase transition \cite{clock_model_dipolar_inf,clock_model_dipolar_nn}. 

\begin{table}
\begin{center}
\caption
{Critical exponents for the continuous phase transition analyzed with classical Monte-Carlo.
\label{table_critical_exponents}}
  \begin{tabular}{| l | c | c | c | c | r |}
    \hline
     & 1/$\nu$ & $\alpha/\nu$ & $\beta/\nu$ &$\gamma/\nu$ &U. Class\\ \hline
   $\theta\approx\pi/5$ & 1&0 & 1/8 & 7/4 & Ising\\ \hline
    $\theta=0$ & 1.05(3) & 0.10(3) & 0.25(2) & 1.5(2)& Unknown \\ \hline
    $\theta=1^0$ & 1.17(3) & 0.32(3) & 0.25(3)& 1.5(3) & Unknown \\
    \hline
  \end{tabular}
\end{center}
\end{table}

A low-temperature phase 
transition of pure dipoles on the kagome lattice was recently observed in the $O(N)$ 
Monte-Carlo studies in the Ref. \onlinecite{tomita_kagome}. The nature of the low-temperature 
spin arrangement was however not resolved due 
to the high computational cost of the $O(N)$ Monte Carlo algorithm  
inversely proportional to temperature.
We have investigated the system at significantly lower 
temperatures, where snapshots of the spin
configurations give
clear evidence of  ferrimagnetic order at $\textbf{k}_{0}=(0,0)$. At the same time, the temperature dependence of the 
static structure factor does not indicate any intermediate ordering between the low-temperature ferrimagnetic state and the high-temperature disordered configuration. Together with our LT and spin-wave studies this rather strongly suggests that ferrimagnetic ${\bf{k}}=(0,0)$ state is the low-temperature configuration of the dipoles.

In the intermediate regime, due to existence of many metastable energy minima, our Monte-Carlo simulations do not equilibrate even for our extensive parallel setup. We thus cannot provide a clear picture of physical quantities  and leave a detailed investigation of this possibly incommensurate regime for future studies.

\section{Discussion and conclusion}

We have determined ground states, excitations, and phase transitions, of  classical Heisenberg spins with exchange and dipolar interactions on the frustrated kagome lattice. 

Our first central result is a determination of the ground state for classical Heisenberg dipoles. This is a ferrimagnetic three-sublattice one. Note that dipolar interactions for Heisenberg spins lead to  ground states in two dimensional systems effectively confined in the plane of the lattice as a result of extensive energy cost of any finite out of plane component \cite{maleev_1976, de_bell_review_thin_films}. In our studies we indeed observe only in-plane spin-states as the ground states of the model. Therefore, the ground states we find also apply to classical XY spins in the plane of the lattice.

Next, we observe that switching on a weak dipolar interaction lifts the extensive ground-state 
degeneracy of the nearest-neighbour model which exists here as it does in many other 
frustrated lattices, e.g. the Archimedean pyrochlore lattice in three dimensions \cite{moessner1998low, moessner1998properties}. 
In both cases, the elementary 
simplices -- triangles for kagome, 
tetrahedra for pyrochlore -- have vanishing total dipole moment in the nearest-neighbour ground 
state; upon adding dipolar interactions, they enter a state where the quadrupole moment of each
simplex also vanishes \cite{palmer_chalker2000order}. However, for stronger values of the dipolar interaction, the suppression of the
leading multipole moment no longer seems to be favourable.
The general principles governing the low-energy states on individual 
clusters \cite{Schonke:2015aa}, 
and how they combine to form a large lattice, is an intriguing topic for future studies.  

The concomitant line of phase transitions into the ferrimagnetic state at dominant $D$ 
appear to have exponents $\nu$ and $\beta$ 
change monotonically with the ratios $\beta/\nu$ and $\gamma/\nu$ constant. This is known in the context of the  "weak universality" hypothesis and often appears in systems with n-fold anisotropy where exponents appear to `drift' from KT values to those of discrete continuous transitions. The presence of an enigmatic slice of the phase diagram where our methods fail to produce a reliable answer further focuses attention on the possibility of the appearance of incommensurate states for delicately 
balanced exchange and dipolar interactions. 

Moreover, it seems rather remarkable that the flat band of zero-energy excitations simply 
moves up in energy without acquiring almost any dispersion. We do note that this phenomenon is not
so uncommon, after all, with a range of different perturbations capable of producing a similar 
phenomenon, a case in point being magnetoelastic interactions\cite{Tchernyshyov:2002aa}. Also, a recent preprint \cite{dAmbrumenil:2015aa} noted the same 
phenomenon for a dipolar magnet on the Gadolinium Gallium Garnet (GGG) lattice, which has historically played an immensely 
important role in the experimental study of frustrated magnetic materials. This may very well be one of the
best experimental handles on dipolar interactions, leading to an almost ${\bf{k}}$-independent resonance in inelastic neutron scattering \cite{sw_maestro_gingras,lee_broholm_kim2000} at a non-zero energy
scaling quite sensitively with the size of the dipolar interaction $\sim\sqrt{D}$. 

The prospect for experimental work in this field is probably better now than it has been for a very long time. 
In an large number of systems the role of dipolar interactions is important or even dominant \cite{spin_ice_monopoles_nature, rozenbaum1991vibrational}. There is significant progress in fabrication of dipolar nano arrays with a complex frustrated lattice geometry \cite{Wang:2006aa, moller2006artificial} as well as recent progress on building a dipolar systems in optical lattices \cite{pupilo_et_al_cold_atoms_review, yao2012topological}. 
In addition recent progress in fabricating thin films of frustrated materials \cite{leusink2014thin,bhuiyan2005growth} suggests a possible route for realization of dipolar films with a kagome geometry. Here the possible candidates for a film realization could be fcc kagome materials $\rm{RhMn_{3}}$, $\rm{PtMn_{3}}$, $\rm{IrMn_{3}}$ \cite{fcc_kagome_films,irmn_long_range_order} where the latter one is commonly used in thin film technology \cite{itmn_films_takahashi1,itmn_films_takahashi2}.  $\rm{RhMn_3}$ and $\rm{PtMn_3}$ have the fcc crystal structure \cite{rhmn_ptmn_structure1,rhmn_ptmn_structure2, rhmn_ptmn_structure3} where magnetic Mn ions reside on the cube faces and the nonmagnetic (Ir) ions site at the cube corners. The magnetic ions can thus be viewed as being on ABC stacked (111) kagome planes, where each site has eight NNs (four in-plane, two to the plane above, and two to the plane below). The (111) plane is perpendicular to the film plane in thin-film applications and thus one deals with a thin stack of $L$ kagome layers. Interestingly the bulk of $\rm{IrMn_3}$ exhibits a long-range magnetic order below $T_{N} \approx 960 K$ \cite{irmn_long_range_order} which is the 3D manifestion of the $120^o$ $q = 0$ spin structure \cite{fcc_kagome_films2} one of the structures found in our studies to be stabilized by weak dipolar interactions. Similar magnetic order is also found in $\rm{RhMn_{3}}$ and $\rm{PtMn_{3}}$. 

We hope that our work will
provide motivation for detailed characterisation of nature and collective behaviour of some of these 
experimental systems.

\acknowledgments

We are grateful to P. Deen, P. A. McClarty,  L. Seabra and R. Valenti for useful discussions. We thank Rechenzentrum Garching (RZG) for computing time for the parallel simulations. M.M. acknowledges ICMP of NAS of Ukraine (Lviv) where part of initial computations was performed. 

\appendix

\section{Linear spin-wave theory}
\label{appendix:spin_waves}
Our spin wave analysis in the non-collinear magnetic
systems starts with a rotation from the global $z$-direction
to the local frame for each moment. Let $\mathbf{\tilde{S}}_{i}(\mathbf{R}^{k})$
point along its local $z$-axis so that it is related to the spin operator
defined in the crystallographic frame via the rotation:
\begin{equation}
\mathbf{S}_{i}(\mathbf{R}^{k})=\mathcal{\mathbb{R}}_{i}^{\text{\textmd{-1}}}\mathbf{\tilde{S}}_{i}(\mathbf{R}^{k})
\end{equation}
where $\mathbb{R}_{i}$ is the corresponding rotation matrix. In
the local frame the Hamiltonian reads
\begin{eqnarray}
H & = & -\frac{1}{2}\sum_{i,j}\sum_{\alpha,\beta}\sum_{k,l}\mathcal{\mathcal{J}}_{ij}^{\alpha\beta}(\mathbf{R}_{ij}^{kl})\tilde{S_{i}^{\alpha}}(\mathbf{R}^{k})\tilde{S^{\beta}}_{j}(\mathbf{R}^{l})
\end{eqnarray}
where the interaction matrix components transform as
\begin{equation}
\mathcal{\mathcal{J}}_{ij}(\mathbf{R}_{ij}^{kl})=\mathcal{\mathbb{R}}_{i}^{\text{\textmd{-1}}}J_{ij}(\mathbf{R}_{ij}^{kl})\mathbb{R}_{j}.
\end{equation}
Fourier transforming  spin operators and interaction
matrix gives
\begin{eqnarray}
\tilde{S_{i}^{\alpha}}(\mathbf{R}^{k}) & = & \frac{1}{\sqrt{N}}\sum_{\mathbf{k}}\tilde{S_{i}^{\alpha}}(\mathbf{k})\;\exp\left[ik\cdot\left(\mathbf{R}^{k}+\mathbf{r}_{i}\right)\right]\\
\mathcal{\mathcal{J}}_{ij}^{ij}(\mathbf{k}) & = & \sum_{kl}\mathcal{\mathcal{J}}_{ij}^{ij}(\mathbf{R}_{ij}^{kl})\exp\left[-i\mathbf{k}\mathbf{\cdot R}_{ij}^{kl}\right]
\end{eqnarray}
where $N$ is the number of underlying Bravais lattice points. Thus, the Hamiltonian in reciprocal space is
\begin{equation}
H=-\frac{1}{2}\sum_{i,j}\sum_{\alpha,\beta}\sum_{\mathbf{k}}\tilde{S_{i}^{\beta}}(\mathbf{k})\mathcal{\mathcal{J}}_{ij}^{ij}(\mathbf{k})\tilde{S^{\beta}}_{j}(-\mathbf{k}).
\end{equation}
The linearized Holstein-Primakoff transformation then gives
\begin{eqnarray}
\tilde{S_{i}^{x}}(\mathbf{k}) & = & \sqrt{\frac{S}{2}}\left[c_{i}^{\dagger}(\mathbf{k})+c_{i}(-\mathbf{k})\right]\\
\tilde{S_{i}^{y}}(\mathbf{k}) & = & i\sqrt{\frac{S}{2}}\left[c_{i}^{\dagger}(\mathbf{k})-c_{i}(-\mathbf{k})\right]\nonumber\\
\tilde{S_{i}^{z}}(\mathbf{k}) & = & \sqrt{N}S\delta_{\mathbf{k},0}\exp\left[-i\mathbf{k}\cdot\mathbf{r}_{i}\right]-\frac{1}{\sqrt{N}}\sum_{\mathbf{k}^{\prime}}c_{i}^{\dagger}(\mathbf{k}^{\prime})c_{i}(\mathbf{k}^{\prime}-\mathbf{k}),\nonumber
\end{eqnarray}
with boson operators  $\left[c_{i}(\mathbf{k}),\; c_{j}^{\dagger}(\mathbf{k}^{\prime})\right]=\delta_{i,j}\delta_{\mathbf{k},\mathbf{k^{\prime}}}$.
Keeping only terms
up to second order, we obtain
\begin{equation}
H=H^{(0)}+H^{(1)}+H^{(2)}
\end{equation}
where
\begin{eqnarray}
H^{(0)} & = & -\frac{1}{2}NS^{2}\sum_{i,j}\mathcal{\mathcal{J}}_{ij}^{zz}(0)\nonumber\\
H^{(1)} & = & -S\sqrt{\frac{NS}{2}}\sum_{i,j}\left[\text{F}_{ij}(0)c_{i}^{\dagger}(0)+\text{F}_{ij}^{\star}(0)c_{i}^{\dagger}(0)\right]\nonumber\\
H^{(2)} & = & -\frac{1}{2}S\sum_{i,j}\sum_{\mathbf{k}}\left[\text{A}_{ij}(\mathbf{k})c_{i}^{\dagger}(\mathbf{k})c_{j}(\mathbf{k})+\text{B}_{ij}(\mathbf{k})c_{i}^{\dagger}(\mathbf{k})c_{j}^{\dagger}(-\mathbf{k})\right.\nonumber\\
 &  & \left.+\text{B}_{ij}^{\star}(\mathbf{k})c_{i}(-\mathbf{k})c_{j}(\mathbf{k})+\text{A}_{ij}^{\star}(\mathbf{k})c_{i}(-\mathbf{k})c_{j}^{\dagger}(-\mathbf{k})\right]
\end{eqnarray}
and 
\begin{eqnarray}
\text{F}_{ij}(0) & = & \mathcal{\mathcal{J}}_{ij}^{xz}(0)+i\mathcal{\mathcal{J}}_{ij}^{yz}(0)\nonumber\\
\text{A}_{ij}(\mathbf{k}) & = & \frac{1}{2}\left\{ \mathcal{\mathcal{J}}_{ij}^{xx}(\mathbf{k})+\mathcal{\mathcal{J}}_{ij}^{yy}(\mathbf{k})-i\left[\mathcal{\mathcal{J}}_{ij}^{xy}(\mathbf{k})-\mathcal{\mathcal{J}}_{ij}^{yx}(\mathbf{k})\right]\right\}
\nonumber\\
&& -\sum_{\gamma}\mathcal{\mathcal{J}}_{i\gamma}^{zz}(0)\delta_{i,j}\\
\text{B}_{ij}(\mathbf{k}) & = & \frac{1}{2}\left\{ \mathcal{\mathcal{J}}_{ij}^{xx}(\mathbf{k})-\mathcal{\mathcal{J}}_{ij}^{yy}(\mathbf{k})+i\left[\mathcal{\mathcal{J}}_{ij}^{xy}(\mathbf{k})+\mathcal{\mathcal{J}}_{ij}^{yx}(\mathbf{k})\right]\right\} .\nonumber
\end{eqnarray}
The equilibrium condition that on every site the effective magnetic field
be parallel to the spin direction implies the absence of linear terms. 
This is satisfied if $\sum_{j}\text{F}_{ij}(0)=0$.
If the spin ground state is stable after the canonical transformation the Hamiltonian can be written in
diagonal form

\begin{eqnarray}
H&=&H^{(0)}+\sum_{\mathbf{k}}\sum_{i}\epsilon_{i}(\mathbf{k})\\
&&+\sum_{\mathbf{k}}\sum_{i}\epsilon_{i}(\mathbf{k})\left[a_{i}^{\dagger}(\mathbf{k})a_{i}(\mathbf{k})+a_{i}^{\dagger}(-\mathbf{k})a_{i}(-\mathbf{k})\right],\nonumber
\end{eqnarray}
where $a_{i}(\mathbf{k})$ and $a^{\dagger}_{i}(\mathbf{k})$ are new boson operators and all the eigenenergies $\epsilon_{i}(\mathbf{k})$ are real.

The specific heat is
\begin{equation}
C_{v}=\frac{\beta^2}{N} \sum_{\mathbf{k}}\sum_{i}\left[\epsilon_{i}(\mathbf{k})n_{\mathrm{B}}(\epsilon_{i}(\mathbf{k}))\right]^{2} \exp\left[\beta\epsilon_{i}(\mathbf{k})\right]
\end{equation}
where $n_{\mathrm{B}}(\epsilon_{i}(\mathbf{k}))=(\epsilon_{i}(\mathbf{k})-1)^{-1}$ is a Bose factor. 
The sublattice magnetization $M(T)$ is obtained by taking into account the role of quantum 
and thermal fluctuations:
\begin{equation}
M(T)=S-\Delta{S}-\frac{1}{N}\sum_{\mathbf{k}}\sum_{i}\left[\mathrm{Q}^{\dagger}\mathrm{Q}\right]_{ii}n_{\mathrm{B}}(\epsilon_{i}(\mathbf{k}))
\end{equation}
where 
\begin{equation}
\Delta{S}=\frac{1}{2}\left(\frac{1}{N}\sum_{\mathbf{k}}\sum_{i}\left[\mathrm{Q}^{\dagger}\mathrm{Q}\right]_{ii}-1\right)
\end{equation}
is the zero-temperature reduction of classical spin polarization and $\mathrm{Q}$ is the matrix 
diagonalizing the spin-wave Hamiltonian.  

\bibliography{dipolarref}

\end{document}